\documentclass[journal,doublecolumn]{IEEEtran}

\IEEEoverridecommandlockouts
\usepackage{cite}
\usepackage{amsthm,amsmath,amssymb}
\usepackage{algorithm}
\usepackage{algorithmic}
\usepackage{setspace}
\usepackage{graphicx}
\usepackage{textcomp}
\usepackage{xcolor}
\usepackage{stfloats}
\usepackage{color}
\usepackage{multirow}
\usepackage{array}
\usepackage{bm}
\usepackage{subfig}
\usepackage{booktabs}
\usepackage{lettrine}

\usepackage{caption}
\newtheorem{Proposition}{Proposition}

\newtheorem{Remark}{Remark}

\usepackage{authblk}
\usepackage[implicit=true]{hyperref} 
\hypersetup{
	colorlinks=true,
	linkcolor=black,
	citecolor=black,
	bookmarksnumbered=false,
	bookmarks=false
}

\begin{document}

\title{Asynchronous MIMO-OFDM Massive Unsourced Random Access with Codeword Collisions \\}
\author{\IEEEauthorblockN{Tianya Li, Yongpeng Wu, \textit{Senior Member, IEEE}, Junyuan Gao, Wenjun Zhang, \textit{Fellow, IEEE}, Xiang-Gen Xia, \textit{Fellow, IEEE}, Derrick Wing Kwan Ng, \textit{Fellow, IEEE}, and Chengshan Xiao, \textit{Fellow, IEEE}}
	\thanks{{Part of this work was presented in IEEE GLOBECOM 2023 \cite{LiGC23}. The work of Y. Wu is supported in part by the National Key R\&D Program of China under Grant 2022YFB2902100, the Fundamental Research Funds for the Central Universities, National Natural Science Foundation (NSFC) under Grant 62122052 and 62071289, 111 project BP0719010, and STCSM 22DZ2229005.
			
			T. Li, Y. Wu (corresponding author), J. Gao, and W. Zhang are with the Department of
			Electronic Engineering at Shanghai Jiao Tong University, Shanghai, China.
			Emails: \{tianya, yongpeng.wu, sunflower0515, zhangwenjun\}@sjtu.edu.cn.
			
			X.-G. Xia is with the Department of Electrical and Computer Engineering, University of Delaware, Newark, DE 19716, USA. E-mail: xianggen@udel.edu.
			
			D. W. K. Ng is with the School of Electrical Engineering and Telecommunications, University of New South Wales, Sydney, NSW 2052, Australia. E-mail: w.k.ng@unsw.edu.au.
			
			C. Xiao is with the Department of Electrical and Computer Engineering, Lehigh University, Bethlehem, PA 18015, USA. E-mail: xiaoc@lehigh.edu.
		}
	}
}

\maketitle
\thispagestyle{empty}

\begin{abstract}
	
	\par This paper investigates asynchronous multiple-input multiple-output (MIMO) massive unsourced random access (URA) in an orthogonal frequency division multiplexing (OFDM) system over frequency-selective fading channels, with the presence of both timing and carrier frequency offsets (TO and CFO) and non-negligible codeword collisions. The proposed coding framework segregates the data into two components, namely, preamble and coding parts, with the former being tree-coded and the latter LDPC-coded. By leveraging the dual sparsity of the equivalent channel across both codeword and delay domains (CD and DD), we develop a message-passing-based sparse Bayesian learning algorithm, combined with belief propagation and mean field, to iteratively estimate DD channel responses, TO, and delay profiles. Furthermore, by jointly leveraging the observations among multiple slots, we establish a novel graph-based algorithm to iteratively separate the superimposed channels and compensate for the phase rotations. Additionally, the proposed algorithm is applied to the flat fading scenario to estimate both TO and CFO, where the channel and offset estimation is enhanced by leveraging the geometric characteristics of the signal constellation. Extensive simulations reveal that the proposed algorithm achieves superior performance and substantial complexity reduction in both channel and offset estimation compared to the codebook enlarging-based counterparts, and enhanced data recovery performances compared to state-of-the-art URA schemes.
	
\end{abstract}


\section{Introduction}	
\par Massive machine-type communications (mMTC), recognized as one of the three pivotal application scenarios in fifth-generation (5G) wireless communications, facilitate connectivity for burgeoning communication services \cite{wu2020wcm}. The two distinct features in mMTC, namely, sporadic traffic patterns and short data payloads, render traditional random access schemes inadequate due to their inefficient handshake procedures. As a promising technology to further improve efficiency, grant-free random access (GF-RA) has garnered remarkable attention for its low signaling overhead and reduced latency \cite{gao2022TIT}. In many literature, GF-RA is commonly referred to as sourced random access (SRA) since it assigns a unique pilot to each user to identify the source of the message\cite{Gao_1,Gao_2}. Nevertheless, assigning unique pilots or encoders to accommodate the vast number of users is deemed impractical. As a prospective paradigm, unsourced random access (URA), first introduced by Polyanskiy in \cite{yury2017isit}, mitigates this challenge by handling massive uncoordinated users through a shared common codebook, where the receiver is tasked to recover messages up to permutations, regardless of user identities.

% TODO 注意此段话的过去时态
\par Initially, a random coding bound for the Gaussian multiple-access channel (GMAC) was discussed in \cite{yury2017isit}. Subsequently, a low-complexity coding scheme combining compute-and-forward and coding for a binary adder channel was investigated in \cite{Ordentlich2017ISIT}, which exhibits an enhanced performance compared to the popular solutions, such as slotted-ALOHA and treating interference as noise (TIN). Nevertheless, the codebook size in \cite{Ordentlich2017ISIT} increases exponentially with the blocklength, rendering prohibitive computational complexity. To mitigate this issue, extensive low-complexity algorithmic approaches have been developed to improve energy and spectral efficiency under GMAC \cite{Amal2020TIT,SPARC2021TIT,Pradhan2019GC,Vem2019TCOM,Pradhan2020ICC,Zheng2020VTC}, fading \cite{Kowshik2019ISIT,yury2019ACSSC,Kowshik2020TCOM}, and multiple-input multiple-output (MIMO) channels \cite{Feng2021TIT,Shao2021GC,Shao2020TSP,Wang2022PIMRC,Shao2023ISIT,li2020GC,li2022JSAC,xie2022TCOM,Shyianov2021JSAC,Gkag2023TCOM,Ahma2024TWC,Che2022JSAC, Zhang2024TWC,Decu2022GC,Decurninge2021WCL}. Generally, these studies can be categorized into three prominent avenues, namely, slotted-based, signature-based, and tensor-based schemes \cite{Che2023Network}. Specifically, the slotted-based scheme employs a divide-and-conquer approach, where both coupled compressed sensing (CCS) and uncoupled CS (UCS) frameworks are developed to partition data into multiple slots to reduce the codebook dimension. In the CCS framework, \cite{Amal2020TIT} introduced a cascaded system, where the inner and outer codes were implemented by the sparse regression code (SPARC) \cite{SPARC2021TIT} and the proposed tree code, respectively. This structure was inherited by the later works in \cite{Feng2021TIT,Shao2020TSP,Wang2022PIMRC,Che2022JSAC}, where the Bayesian or covariance-based methods were leveraged to recover the data embedded in each slot. Opposite to CCS, to improve spectral efficiency, no outer coding technique is adopted in the UCS framework, where the data is spliced by clustering the spatial domain correlated channels \cite{xie2022TCOM, Shyianov2021JSAC}. On the other hand, the signature-based scheme, which has demonstrated superiority over its slotted counterpart, has gained widespread adoption for its innovative data division into preamble and coding sections \cite{Vem2019TCOM}. The retrieval of the data embedded in preambles is formulated to the joint activity detection and channel estimation (JADCE) problem. Targeted to this problem, various methodologies, such as message passing (MP) \cite{jiang2023TCOM}, multiple measurement vector approximate MP (MMV-AMP) \cite{Liu2018tsp,Wang2023TWC}, generalized MMV-AMP (GMMV-AMP) \cite{ke2020tps}, and sparse Bayesian learning (SBL) \cite{Zhang2018TVT} algorithms exhibit satisfying performance. The remaining data in the coding part is encoded by the forward-error-correction (FEC) codes, such as the low-density parity-check (LDPC) codes \cite{Pradhan2019GC,Zhang2024TWC,li2020GC,li2022JSAC}, and the polar codes\cite{Pradhan2020ICC,Zheng2020VTC,Ahma2024TWC}. Different from the above two schemes, the tensor-based scheme is a non-segmented scheme, where the FEC-coded messages are modulated and mapped as ranked-one tensors, reducing the computational demands for multi-user separation \cite{Shao2021GC,Decu2022GC,Decurninge2021WCL}. In the literature, the polar coding combined with the random spreading modulation, leading to the fading spread URA (FASURA) \cite{Gkag2023TCOM}, approaches the bound derived in \cite{Gao2023WCL} and has emerged as the state-of-the-art MIMO-URA solution.

\par In practice, since a codebook is shared among massive numbers of users, codeword collisions inevitably occur when multiple users select the same codeword within a time slot, which results in decoding failures and becomes the bottleneck in URA \cite{li2022JSAC}. To address this critical issue, numerous studies have opted to enlarge the codebook size to reduce the collision probability and neglect the collisions for simplicity \cite{Gkag2023TCOM, Vem2019TCOM, Shyianov2021JSAC}. Nonetheless, the unmanageable dimension (e.g., $2^{16}$) imposes substantial computational demands and is prohibitive for memory-limited URA applications employing low-cost sensors. In the literature, there are also some methods resolving collisions while maintaining the codebook dimension moderate. For example, \cite{xie2022TCOM} separated the superimposed channels by exploiting the diversity on the virtual angle domain, although it is only effective with a very limited number of collided users. Also, a window-sliding protocol was established in \cite{li2022JSAC} to address collisions by data retransmission, albeit at the expense of latency and efficiency. Besides, multi-stage orthogonal pilots were utilized in \cite{Ahma2024TWC} to reduce collision density, where the data is decoded based on the individually estimated channel in each slot, leading to increased complexity and CE suboptimality. Despite various efforts have been devoted, the aforementioned solutions may be generally unsatisfactory either in terms of complexity, efficiency, or collision resolution capabilities.

\par Furthermore, due to the inherent simplicity of the one-step RA procedure and the absence of synchronization mechanisms, users may access the base station (BS) asynchronously. In reality, due to heterogeneous transmission distances and oscillator variations, the received signals are affected by both timing and carrier frequency offsets (TO and CFO), inducing random phase rotations in frequency and time domains (FD and TD), respectively, which will cause significant performance degradation if not well compensated. Consequently, various advanced strategies, including TIN \cite{yury2019ACSSC}, CS \cite{Amal2019ICASSP}, orthogonal AMP (OAMP) \cite{Bian2023GC},  tensor-based modulation (TBM) \cite{Decu2022GC}, and SBL \cite{Salari2021CL} are leveraged to deal with TO or CFO in URA scenarios. Additionally, numerous studies have also been dedicated to exploring asynchronous (Async) configurations in GF-RA systems. For example, based on the formulated JADCE problem, \cite{sun2022TWC} developed the structured generalized AMP (S-GAMP) algorithm to jointly estimate TO and CFO in MIMO-orthogonal frequency division multiplexing (OFDM) systems, and \cite{Bian2023GC} utilized the memory AMP algorithm to estimate TO with a reduced computational complexity. \cite{Zhang2023IOT} employed the SBL algorithm to mitigate the inter-symbol interference (ISI) caused by TO, albeit under the assumption of a perfectly known delay profile for user differentiation. Besides, \cite{Guo2024IOTJ} employed the chirp signals to address TO by leveraging the auto-correlation feature. Note that \cite{sun2022TWC, Amal2019ICASSP,Bian2023GC} own a common feature that the pilot matrix is expanded by quantizing all possible TOs, which sharply increases the computational complexity. Besides, \cite{sun2022TWC, yury2019ACSSC, Decu2022GC} utilized OFDM technology yet solely considered the flat fading channel. However, in practice, the complicated scattering environment induces multipath effect and frequency-selective fading (FSF), which leads to the channel varying on subcarriers and renders the aforementioned works inapplicable.  To cope with this issue, various techniques, such as turbo MP \cite{Jiang2022TWC}, SBL \cite{Zhu2022TCOM, Salari2021CL}, and orthogonal matching pursuit (OMP) \cite{URA_FS} are leveraged to address the FSF and retrieve channel responses.

% TODO 在窄带系统中，占用的子载波较少时，可以认为信道是平坦的

\par From the above state-of-the-art overview, it is evident that existing researches primarily focus on URA, either in the synchronous (Sync) or Async scenarios with only TO or CFO and the assumption of no collision. However, the presence of both TO and CFO is inevitable due to the transmission distance and oscillator variations. In such cases, the challenge lies in that the received signal will be coupled with phase rotations in both FD and TD, rendering existing methodologies incompetent \cite{yury2019ACSSC, Amal2019ICASSP, Decu2022GC, Salari2021CL}. Besides, the estimated channel will also be affected by the phase rotations, leading to the channel-based collision resolution schemes not compatible \cite{xie2022TCOM, li2022JSAC, Ahma2024TWC}. Although the number of collisions can be reduced to a certain extent by enlarging codebook dimensions, it is at the expense of significant complexity. Furthermore, when applying OFDM technology to manage TO, most works focus on the flat fading channel. However, due to the complicated scattering environment in practical scenarios, the inherent FSF nature of the channel introduces significant challenges for CE due to the enlarged dimension of unknown channel responses. Motivated by the above considerations, for the first time, we attempt to achieve reliable communications in an Async MIMO-OFDM URA system in the presence of both TO and CFO over FSF channels, while effectively resolving codeword collisions. The main contributions are summarized as follows:
\begin{itemize}
	\item We consider an Async OFDM transmission framework with channel frequency-selectivity in the URA system. By leveraging the dual sparsity of the FSF channel in both codeword and delay domains (CD and DD), we utilize the low-complexity \textit{MP-based SBL algorithm for JADCE (JADCE-MP-SBL)} to jointly estimate DD channel responses and delay profiles integrated with TO. Combined with belief propagation (BP) and mean-field (MF) algorithms, the channel and underlying hyper-parameters are iteratively updated by message calculations, with the complexity linear to the codebook and antenna sizes, making it computationally efficient for large codebooks and massive MIMO settings.
	
	\item Considering that the estimated channel is coupled with CFO-caused phase rotations and superimposed due to collisions, we propose a novel \textit{graph-based channel reconstruction and collision resolution (GB-CR$^2$)} algorithm to iteratively reconstruct channels, resolve collisions, and compensate for quantized CFOs jointly among multiple slots. By leveraging channel information along time slots, the proposed algorithm notably reduces the number of erroneous paths (EPs) compared to the conventional tree code method \cite{Amal2020TIT}, and exhibits a substantial computational complexity reduction in offset estimation compared to the codebook enlarging-based methodologies \cite{sun2022TWC, Amal2019ICASSP}.
% {\bl Specifically, we construct a bipartite graph model and prioritize the most credible path to iteratively retrieve channels and estimate CFOs.}

	\item In the scenario of a narrowband OFDM system with only a small fraction of subcarriers occupied, the channel can be modeled as frequency-flat. In this context, we further apply the GB-CR$^2$ algorithm to this special case to illustrate its effectiveness and versatility. By adopting the orthogonal codebook, the coupled phase rotations are shifted from codewords to channels, thus avoiding the need to enlarge the codebook due to offset quantization. To cope with the accuracy degradation due to the estimation of both TO and CFO, we innovatively leverage the geometric characteristics of the signal constellation to enhance the offset and channel estimation results.
\end{itemize}

\par In the rest of this paper, Section \ref{sec-2} presents the system model. The proposed scheme including the encoding and receiver designs is detailed in Section \ref{sec-3}. Section \ref{sec-4} discusses the application of the propose algorithm to the flat fading channel scenario. Performance analysis and extensive numerical results are provided in Sections \ref{sec-5} and \ref{sec-VI}, respectively. Finally, Section \ref{sec-7} concludes the paper.

\par \textit{Notations:} Throughout this paper, scalars, vectors, and matrices are denoted by lowercase, boldface lowercase, and boldface uppercase, respectively. The transpose, conjugate, and conjugate transpose operations are denoted by $\left( \cdot\right)^T, \left( \cdot\right)^*, \left( \cdot\right)^H$, respectively. $\left\| \mathbf{x} \right\|_p$ and $\left\| \mathbf{A} \right\|_F$ denote the standard $l_p$ and Frobenius norms, respectively. $\text{diag} \left\lbrace \mathbf{d} \right\rbrace$ denotes a diagonal matrix with the vector $\mathbf{d}$ being the diagonals. $\left[T\right]$ denotes the set of integers from $1$ to $T$. $|\mathcal{X}|$ denotes cardinality of  the set $\mathcal{X}$. $\mathbf{A}_{[\mathbf{m},:]}, \mathbf{A}_{[:,\mathbf{n}]}, \mathbf{A}_{[\mathbf{m},\mathbf{n}]}$ represent the sub matrices by extracting the rows indexed by $\mathbf{m}$, the columns indexed by $\mathbf{n}$, and the both, respectively. $\mathbf{A}(m,:), \mathbf{A}(:,n), \mathbf{A}(m,n)$ denote the $m$-th row, $n$-th column, and the $(m,n)$-th entry of $\mathbf{A}$, respectively. $\mathcal{CN}(x;\mu,\sigma^2)$ denotes the complex Gaussian distribution of a random variable $x$ with mean $\mu$ and variance $\sigma^2$. The expectation w.r.t. a function $g(x)$ is denoted by $\left\langle f(x) \right\rangle_{g(x)} = \int f(x)g(x) dx \slash \int g(y) dy$. $\text{Tr}(\cdot)$ denotes the trace of a matrix. $\mathbf{I}_N$ represents the $N \times N$ unit matrix.  $\odot$ and $\otimes $ denote the element-wise multiplication and convolution, respectively. $\mathbb{C}$ and $\mathbb{Z}$ denote the complex and integer fields, respectively. $\propto$ denotes the direct proportionality. $\chi_{n}^2$ denotes the chi-square distribution with $n$ degrees of freedom.
	
\section{System Model} \label{sec-2}

\par Consider $K_a$ active users in a single-cell MIMO cellular network served by a BS equipped with $M$ antennas in an unsourced manner. Namely, no pilot resource is allocated in advance and users select codewords from a shared codebook to access the BS. More generally, we consider an asynchronous transmission via a multipath fading channel in this paper. For each user $k\in [K_a]$, let $\tau_k, \epsilon_k, L_k, \alpha_{k,l,m}$, and $\tau_{k,l}$  denote the normalized TO and CFO, number of channel taps, channel gain of the $l$-th tap and the $m$-th antenna, and normalized delay of the $l$-th tap, respectively, where $\alpha_{k,l,m}  \sim \mathcal{CN}(0, \sigma_h^2)$ with $ \sigma_h^2$ being the energy of the channel tap. 

\par As a mature technical solution, OFDM is employed to combat the multipath effect. An OFDM symbol contains $N_c$ orthogonal  subcarriers, $S$ of which are employed for transmission and the rest are for random data. 
Let $\Delta f$ denote the subcarrier spacing and $W \triangleq N_c \Delta f$ is the total transmission bandwidth. A cyclic prefix (CP) with length $L$ is inserted in the transmitted symbol as a safeguard. Denote the FD data of user $k$ at the $t$-th OFDM symbol as ${\mathbf{x}}_k^t\in\mathbb{C}^{S\times1}$, which is modulated to the TD signal as $\tilde{\mathbf{x}}_k^t = \mathbf{F}_{[\mathbf{s},:]} ^H {\mathbf{x}}_k^t \in\mathbb{C}^{N_c\times1}$, where $ \mathbf{F}_{[\mathbf{s},:]}  \in\mathbb{C}^{S\times N_c}$ is the partial $N_c$-point discrete Fourier transformation matrix $\mathbf{F} \in \mathbb{C}^{N_c\times N_c}$ with $\mathbf{F}(n,k)  = \frac{1}{\sqrt{N_c}}e^{-j2\pi nk\slash N_c}$, and $\mathbf{s}=[n_1,\cdots,n_S]^T \in \mathbb{Z}^{S\times 1}$ is the vector of the occupied subcarrier indices. After removing the CP sequences, the TD received signal $\tilde{\mathbf{y}}_m^t \in \mathbb{C}^{N_c \times 1}$ at the $m$-th antenna on the $t$-th symbol is given by
\begin{equation}
		\tilde{\mathbf{y}}_m^t  =  \sum\limits_{k=1}^{K_a}{ \mathbf{D}_{\epsilon_k}^t  \left(  \mathbf{I}_{N_c}\right) _{\tau_k} \tilde{\mathbf{x}}_k^t   \otimes   \tilde{\mathbf{h}}_{k,m}  }  + \tilde{\mathbf{z}},  \label{equ-1}
\end{equation} 
where $\mathbf{D}_{\epsilon_k}^t =\phi^t \text{diag} \left\lbrace  (1, \omega_k,\cdots,\omega_k^{N_c-1}) \right\rbrace \in\mathbb{C}^{N_c\times N_c}$ denotes the phase shift caused by $\epsilon_k$, with $\omega_k = e^{{j2\pi \epsilon_k}\slash{N_c}}$; $\phi^t_k = \omega_k^{L+(t-1)(L+N_c)}$ represents the phase shift accumulated to the $t$-th symbol; $\left(\mathbf{I}_{N_c}\right)_{\tau_k}$ denotes the unit matrix left-cyclically shifted with $\tau_k$ units; $\tilde{\mathbf{h}}_{k,m} = \big[\tilde{h}_{k,m}[0], \cdots, \tilde{h}_{k,m}[L_k-1] \big]^T \in\mathbb{C}^{L_k\times 1}$ denotes the channel taps of user $k$ on the $m$-th antenna and $\tilde{\mathbf{z}} \in\mathbf{C}^{N_c\times 1}$ represents the additive white Gaussian noise (AWGN) with each component distributed as $\mathcal{CN}\left(0, \sigma_n^2 \right)$. Note that to obtain Eq. \eqref{equ-1}, we assume that $L > \tau_{k,l}+\tau_{k}, \forall k\in [K_a], l\in [L_k]$ such that there is no ISI among OFDM symbols. The FD-demodulated signal ${\mathbf{y}}_m^t \in \mathbb{C}^{S\times 1}$ is given by
\begin{equation}
	\begin{aligned}
		{\mathbf{y}}_m^t  &= \mathbf{F}_{[\mathbf{s},:]}  \tilde{\mathbf{y}}_m^t  \\
		&=  \sum\limits_{k=1}^{K_a} { \mathbf{F}_{[\mathbf{s},:]}  \mathbf{D}_{\epsilon_k}^t    \left(   \mathbf{I}_{N_c}\right) _{\tau_k}\mathbf{F}_{[\mathbf{s},:]} ^H {\mathbf{x}}_k^t} \odot {\mathbf{h}}_{k,m} + {\mathbf{z}}\\
		&\triangleq \sum\limits_{k=1}^{K_a} {  \mathbf{P}_{\epsilon_k}^t  \mathbf{P}_{\tau_k} {\mathbf{X}}_k^t}{\mathbf{h}}_{k,m} +{\mathbf{z}},
	\end{aligned}  \label{equ-2}
\end{equation}
where $\mathbf{X}_k^t \triangleq \text{diag} \left\lbrace \mathbf{x}_k^t \right\rbrace \in\mathbb{C}^{S\times S}, \mathbf{z} \triangleq \mathbf{F}_{[\mathbf{s},:]} \tilde{\mathbf{z}} \in \mathbb{C}^{S\times 1}$, and ${\mathbf{h}}_{k,m} \triangleq  \mathbf{F}_{[\mathbf{s},1:L]}  \tilde{\mathbf{h}}_{k,m} \in \mathbb{C}^{S\times 1}$ is the FD channel,  which reads
\begin{equation}
	{\mathbf{h}}_{k,m}(s) = \sum\limits_{l=0}^{L_k-1} \alpha_{k,l,m} e^{\frac{-j2\pi(n_s-1) }{N_c}{\tau}_{k,l}}, s \in [S].   \label{equ-4}
\end{equation}
The matrices $\mathbf{P}_{\epsilon_k}^t, \mathbf{P}_{\tau_k} \in \mathbb{C}^{S\times S}$ in Eq. \eqref{equ-2} denote the phase rotation matrices caused by the CFO $\epsilon_k$ and TO $\tau_k$, respectively, which are given by
\begin{subequations}
	\begin{align}
		\mathbf{P}_{\epsilon_k}^t &=   \mathbf{F}_{[\mathbf{s},:]}   \mathbf{D}_{\epsilon_k}^t \mathbf{F}_{[\mathbf{s},:]} ^H = 	\phi^t_k \mathbf{P} _{[\mathbf{s},\mathbf{s}]}, \label{equ-5a} \\
		\mathbf{P}_{\tau_k} &  = \mathbf{F}_{[\mathbf{s},:]}   \left(\mathbf{I}_{N_c}\right) _{\tau_k} \mathbf{F}_{[\mathbf{s},:]} ^H = \text{diag} \left\lbrace \bm{\psi}_k \right\rbrace,  \label{equ-5b}
	\end{align}  
\end{subequations}
where $\bm{\psi}_k = [ \psi_k^{1-n_1} \!, \!\cdots\!,\! \psi_k^{1-n_S} ]^T \! \in \! \mathbb{C}^{S\times 1}$ with $\psi_k = e^{j2\pi \tau_k \slash N_c}$. $\mathbf{P}\in\mathbb{C}^{N_c\times N_c}$ is given by
\begin{equation}
	\mathbf{P} = \left[\!\!\!              %左括号
	\begin{array}{cccc}   %该矩阵一共3列，每一列都居中放置
		P(\epsilon_k) & P(1+\epsilon_k) & \cdots & P(N_c \!-\!1 \!+\!  \epsilon_k)\\  %第一行元素
		P(N_c \!-\!1 \!+\!  \epsilon_k) &  P(\epsilon_k) & \cdots & P(N_c \!-\! 2 \!+\! \epsilon_k)\\  %第二行元素
		\vdots &  \vdots &  \ddots &  \vdots \\
		P(1\!+\! \epsilon_k) & P(2\!+\!\epsilon_k)  & \cdots & P(\epsilon_k)
	\end{array} \!\!\!\right], 
\end{equation}
where 
\begin{equation}
	P(\epsilon_k) = \frac{\sin \pi \epsilon_k}{N_c \sin \pi \epsilon_k\slash N_c} e^ {j  \pi \epsilon_k (N_c-1)\slash N_c }. 
\end{equation}
The matrix $\mathbf{P}$ indicates that the CFO induces the inter-carrier interference (ICI) and associated attenuation of the received signal since $\left| P(\epsilon_k^t)\right| < 1$ when $\epsilon_k \neq 0$. In practical systems such LTE or 5G NR, users are required to estimate frequency offsets by detecting the downlink synchronization signals during the cell search procedure and utilize the frequency offset compensation methods to compensate for the large frequency offsets  \cite{sun2022TWC}. This closed-loop synchronization mechanism effectively maintains the residual CFO within a small range \cite{CFO2014TWC,CFO2008SPL}. Therefore, for the sake of model tractability and algorithm illustration, we consider the range of CFO within $\pm 200$ Hz, with the normalized value within $\pm 0.013$ for a subcarrier spacing $\Delta f = 15$ kHz. Thus, the ICI is negligible, and $\mathbf{P}_{\epsilon_k}^t$ is assumed to be diagonal. Then, we employ a vector $\mathbf{p}_k^t \in \mathbb{C}^{S\times 1}$ to approximate the diagonals of the phase  rotation matrix, i.e., 
\begin{equation}
	\mathbf{p}_k^t(s) =  \omega_k^{(L+N_c)t- {(N_c+1)}\slash{2}} \psi_k^{1-n_s}, s\in [S].  \label{equ-6}
\end{equation}
Note that neglecting ICI does not imply that the CFO-caused phase rotation can be also ignored. In fact, the phase shift is accumulates over the sampling time due to the factor $\phi_k^t$, resulting in a non-negligible phase rotation even for small CFO values.

\begin{figure*}[t]
	%	\vspace{-0.2cm}
	\centerline{\includegraphics[width=0.95\textwidth]{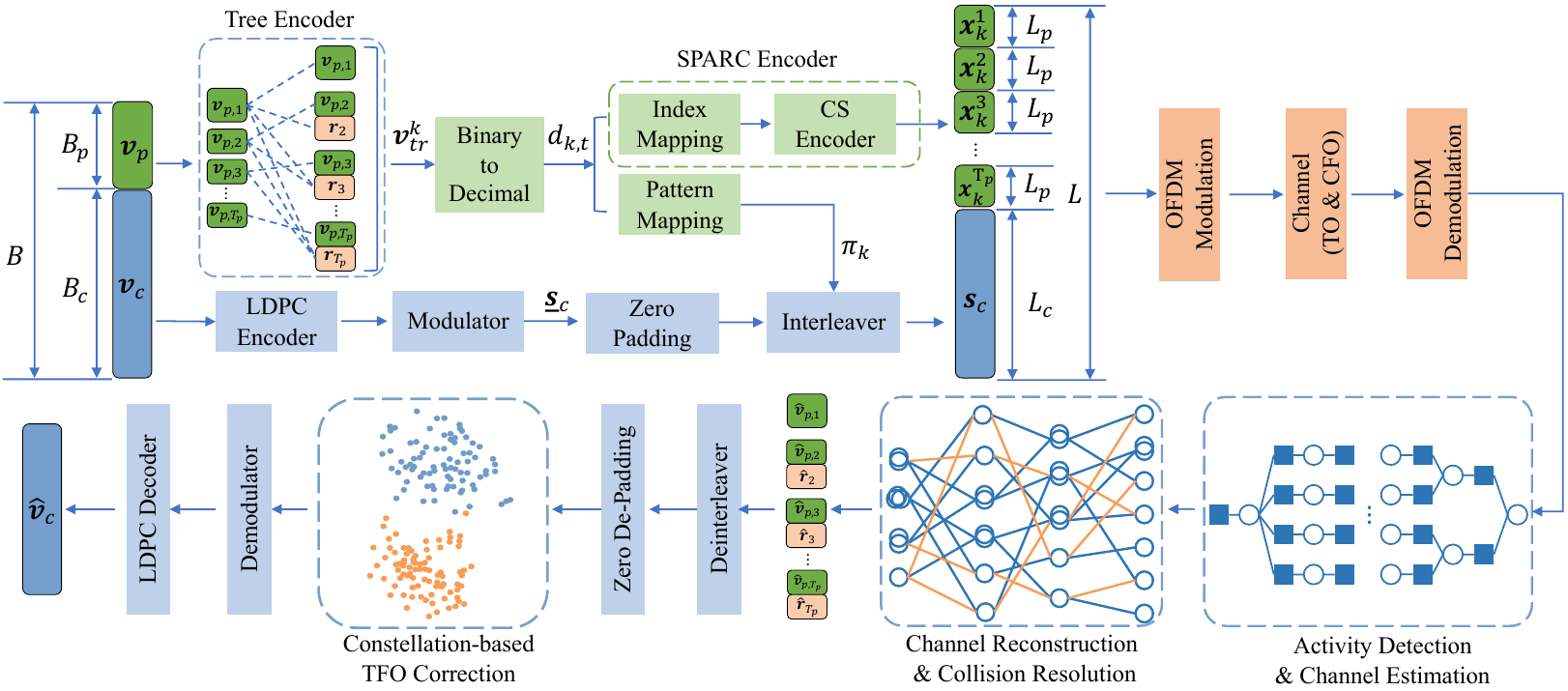}}
	\caption{A realization of the overall encoding scheme and the proposed receiver design.}
	\label{pic-1}
	%\vspace{-0.5cm}
\end{figure*}

\section{Proposed Scheme} \label{sec-3}
\par This section presents the overall encoding scheme and receiver design, visually depicted in Fig. \ref{pic-1}. The focal point is the detailed discussion of the proposed JADCE-MP-SBL and GB-CR$^2$ algorithms.

\subsection{Encoding Scheme} \label{sec-3-1}
\par The user's information is partitioned into two parts, with $ \mathbf{v}_p \in \left\lbrace 0,1 \right\rbrace^{B_p\times1} $ being the first part and $\mathbf{v}_c\in \left\lbrace 0,1 \right\rbrace^{B_c\times1}$ the second part, referred to as the preamble and coding parts, each occupied $T_p$ and $T_c$ OFDM symbols, respectively. In most literature, the preamble part handles tasks such as CE and recovery of key parameters such as interleaving patterns \cite{li2022JSAC, Vem2019TCOM} or spreading sequences \cite{Gkag2023TCOM}. In this paper, in addition to the basic tasks mentioned above, the preamble part takes on two more crucial tasks: 1) codeword collision resolution and 2) TO and CFO estimation. Firstly, the preamble sequence $\mathbf{v}_p$ is partitioned into $T_p$ sub-blocks, i.e., $\mathbf{v}_p = [\mathbf{v}^T_{p,1}, \cdots, \mathbf{v}^T_{p,T_p}]^T$ with $\mathbf{v}_{p,t} \in \{0,1\}^{b_t \times 1}$ and $\sum\nolimits_{t=1}^{T_p}b_t = B_p$. Then, the tree encoder generates parity bits appended to the preamble to facilitate data splicing  \cite{Amal2020TIT}. Specifically, each sub-block is resized to length $J$ by appending $p_t = J - b_t$ parity bits for $t\in [2:T_p]$ and $b_1=J$. Hence, the tree-coded message is $\mathbf{v}_{\mathrm{tr}} = [\mathbf{v}^T_{\mathrm{tr},1},\cdots, \mathbf{v}^T_{\mathrm{tr},T_p}]^T  \in \{0,1\}^{ JT_p \times 1}$ with $\mathbf{v}_{\mathrm{tr},1}=\mathbf{v}_{p,1}$ and  $\mathbf{v}_{\mathrm{tr},t} = [ \mathbf{v}^T_{p,t}, \mathbf{r}^T_t]^T$ for $t\in [2:T_p]$. The parity bit vector $\mathbf{r}_t \in \{0,1\}^{p_t\times 1}$ is  given by
	\begin{equation}
		\mathbf{r}_t= \sum\nolimits_{s=1}^{t-1}  \mathbf{G}_{s,t-1} \mathbf{v}_{p,s}, t\in [2:T_p],  \label{equ-7}
\end{equation}
where $\mathbf{G}_{s,t-1}\in \{0,1\}^{p_t\times b_t}$ is a random binary matrix with the entries independent Bernoulli trials. The arithmetics in Eq. \eqref{equ-7} are modulo-2 and, as such, $\mathbf{r}_t$ remains binary. For each active user $k$, the coded message $\mathbf{v}_{\mathrm{tr}}^k = [(\mathbf{v}_{\mathrm{tr},1}^k)^T, \cdots, (\mathbf{v}_{\mathrm{tr},T_p}^k)^T]^T$ is then mapped to the decimal number $d_{k,t} = [\mathbf{v}_{\mathrm{tr},t}^k]_{10}+1$. In the SPARC encoder, each user picks the codeword from the codebook $\mathbf{A} = \left[\mathbf{a}_1, \mathbf{a}_2, \cdots, \mathbf{a}_{N} \right] \in \mathbb{C}^{L_p \times N}$ with $\left\| \mathbf{a}_n \right\| ^2 = L_p$, $N=2^J$, and $L_p = S$. Specifically, by mapping the decimal message $d_{k,t}$ to the index of the codeword, user $k$ selects $\mathbf{a}_{d_{k,t}}$ as the transmitted message in the $t$-th slot. With a slight abuse of notation, we define $\mathbf{c}_k^t \in \{0,1\}^{N \times 1}$ as a binary selection vector, which is all-zero but a single one in position $d_{k,t}$. Thus, reviewing Eq. \eqref{equ-2}, we have $\mathbf{x}_k^t = \mathbf{A}\mathbf{c}_k^t, t \in[T_p]$, corresponding to $T_p$ OFDM symbols. While in the coding part, the interleaving pattern $\pi_k$ is codetermined by $\{d_{k,t}\}_{t=1}^{T_p}$, i.e., $\pi_k = f (d_{k,1},\cdots, d_{k,T_p})$ for some mapping function $f(\cdot)$. The message $\mathbf{v}_c$ is LDPC-coded, modulated, zero-padded and interleaved to $\mathbf{s}_c\in \left\lbrace 0,1 \right\rbrace^{L_c\times1}$ sequentially, which is the so-called interleave-division multiple access (IDMA) \cite{ping2006twc}. Finally, the encoded message is $[(\mathbf{x}_k^1)^T, \cdots, (\mathbf{x}_k^{T_p})^T, \mathbf{s}_c^T]^T \in \mathbb{C}^{ L_{\mathrm{tot}}\times 1}$, corresponding to $T_p+T_c$ OFDM symbols with $L_c = ST_c $ and $L_{\mathrm{tot}} = ST_p+ L_c$.

\subsection{JADCE-MP-SBL Algorithm}
\par
\begin{figure}[htpb]
	\centerline{\includegraphics[width=0.38\textwidth]{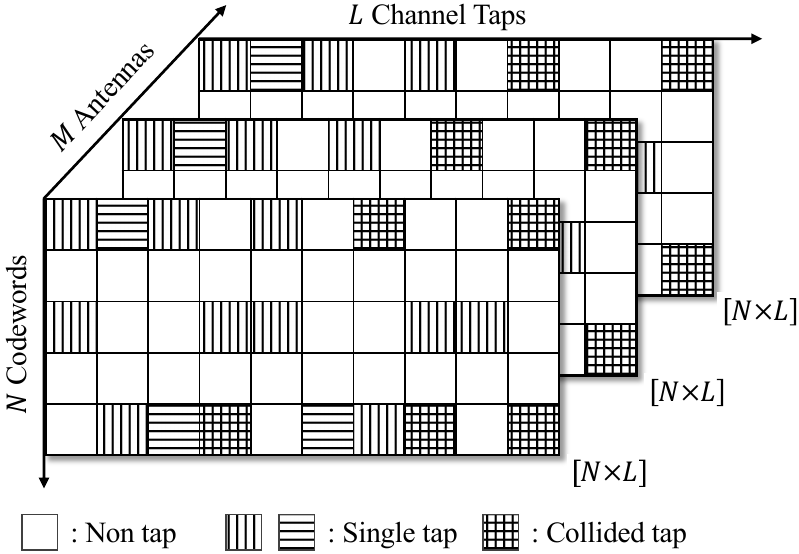}}
	\caption{Demonstration of the sparsity of the matrix $\mathbf{X}^t$.}
	\label{pic-4}
\end{figure}

 Regarding the approximate diagonal properties of the phase rotation, we can rewrite Eq. \eqref{equ-2} into the matrix form as below
\begin{equation}
	\begin{aligned}
		\mathbf{Y}^t &= \sum_{k=1}^{K_a}{p_k^t \text{diag}\left\lbrace \mathbf{A}\mathbf{c}_k^t \right\rbrace \mathbf{H}_k + \mathbf{Z}}\\
							   &= \sum_{k=1}^{K_a}{p_k^t \text{diag}\left\lbrace \mathbf{A}\mathbf{c}_k^t \right\rbrace \mathbf{F}_{[\mathbf{s},1:L]} \tilde{\mathbf{H}}_k + \mathbf{Z}}\\
							   &= \mathbf{G}\tilde{\mathbf{C}}^t \tilde{\mathbf{P}}^t \tilde{\mathbf{H}} + \mathbf{Z}, t\in [T_p],
	\end{aligned} \label{equ-8}
\end{equation}
where $\mathbf{Y}^t \in \mathbb{C}^{S\times M}, \mathbf{H}_k \in \mathbb{C}^{S\times M}$ and $\mathbf{Z}\in\mathbb{C}^{S\times M}$ is the AWGN.  $\mathbf{G} = \left[\mathbf{G}_1,\cdots, \mathbf{G}_L \right]\in \mathbb{C}^{S\times NL}$ is the equivalent sensing matrix with $\mathbf{G}_l = \left[\mathbf{a}_1\odot \mathbf{f}_l, \cdots, \mathbf{a}_N \odot \mathbf{f}_l \right] \in\mathbb{C}^{S\times N}$, where $\mathbf{f}_l \in \mathbb{C}^{S\times 1} $ is the $l$-th column of  $\mathbf{F}_{[\mathbf{s},1:L]}$. $\tilde{\mathbf{C}}^t = \text{diag} \left\lbrace \mathbf{C}^t, \cdots, \mathbf{C}^t \right\rbrace \in \left\lbrace 0,1 \right\rbrace ^{NL\times K_aL}$ is a block-diagonal matrix, where $\mathbf{C}^t = \left[ \mathbf{c}_1^t, \cdots, \mathbf{c}_{K_a}^t\right] \in \left\lbrace 0,1 \right\rbrace ^{N\times K_a}$ is the user selection matrix in the $t$-th symbol. Generally, the codebook with a lager size is preferred in most literature \cite{Gkag2023TCOM,Amal2020TIT,Feng2021TIT,Shyianov2021JSAC,xie2022TCOM,li2022JSAC,Vem2019TCOM}, i.e., $N=2^{12} \sim 2^{17}$. While the number $K_a$ of active users is usually dozens to hundreds. Therefore, the user selection matrix $\mathbf{C}^t$ is ultra-sparse, where only the $\{(d_{k,t}, k)\}_{k=1}^{K_a}$-th entries are non-zero (ones) and the rest entries are all zeros. We further define $c_ n \triangleq \sum\nolimits_k \mathbf{c}_k^t(n)$, which can be $0, 1$ or more than one, referring to as no user, single user, or multiple users select the $n$-th codeword. For the situation of $c_n > 1$, we say that the codeword collision occurs at $\mathbf{a}_n$. $\tilde{\mathbf{P}}^t = \text{diag} \left\lbrace \mathbf{P}^t, \cdots, \mathbf{P}^t \right\rbrace \in \mathbb{C} ^{K_aL\times K_aL}$ is also a block-diagonal matrix, where $\mathbf{P}^t = \text{diag} \left\lbrace p_1^t, \cdots, p_{K_a}^t \right\rbrace \in \mathbb{C}^{K_a \times K_a}$ is the CFO rotation matrix with $p_k^t \triangleq \phi_k^t P(\epsilon_k)$. Note that TO has been merged into $ \mathbf{H}_k$ such that 
\begin{equation}
	\mathbf{H}_k(s,m) = \sum\limits_{l=0}^{L_k-1} {\alpha_{k,l,m} e^{\frac{-j2\pi(n_s-1)}{N_c} ({\tau}_{k,l} + {\tau}_k) }}. \label{equ-9}
\end{equation}
 We denote $I_{k,l} \triangleq  {\tau}_{k,l} + {\tau}_k $ as the equivalent $l$-th channel tap of user $k$, which is assumed to be integer-sampled since the resolution of the transmission bandwidth $W$ is sufficient \cite{URA_FS}. On such bases, $\tilde{\mathbf{H}}_k \in \mathbb{C}^{L\times M}$, user $k$'s DD channel, is row-sparse and consists of $L_k$ non-zero rows of the indices $\{I_{k,l}\}_{l=1}^{L_k}$. The channel matrix $\tilde{\mathbf{H}} = [\tilde{\mathbf{H}}_1^T, \cdots, \tilde{\mathbf{H}}_L^T]^T \in\mathbb{C}^{K_aL \times M}$, where $\tilde{\mathbf{H}}_l \in\mathbb{C}^{K_a \times M}$ corresponds to the channel of the $l$-th tap. Let $\mathbf{X}^t \triangleq \tilde{\mathbf{C}}^t \tilde{\mathbf{P}}^t \tilde{\mathbf{H}} \in \mathbb{C}^{NL\times M}$ denote the equivalent channel matrix, of which the pictorial representation is given in Fig. \ref{pic-4}. For clarity, we reshape $\mathbf{X}^t$ to a 3D $N\times L\times M$ tensor. Recall that $\tilde{\mathbf{C}}$ is an ultra-sparse matrix, and $\mathbf{X}^t$ is thus row-sparse along the codeword axis, which exhibits the CD sparsity. Besides, $\mathbf{X}^t$ is intra-row sparse on the tap axis due to the limited number of channel taps, which is the so-called DD sparsity. Also, the dual sparsity on both CD and DD is  duplicated along the antenna axis. In the presence of codeword collisions, the channel taps of multiple users will be mixed. The codeword collisions will be addressed in Sec. \ref{sec-IV-B}, and we focus on the sparse recovery of $\mathbf{X}^t$ here. 
\par We follow the Bayesian approach to retrieve the sparse matrix $\mathbf{X}^t$ from the received noisy superposition. For clarity, we omit the superscript $t$ in all involved variables. The prior probability density function (pdf) for $\mathbf{X}$ is assumed to be  a two-hierarchical structure, i.e., $p(\mathbf{X}) = \int_{\boldsymbol{\gamma}} {p(\mathbf{X} | \boldsymbol{\gamma} ) p(\boldsymbol{\gamma})}$, where $p(\mathbf{X} | \boldsymbol{\gamma} )$ is the conditional prior pdf on $\boldsymbol{\gamma}$ and $p(\boldsymbol{\gamma})$ is a hyper-prior pdf of the precision $\boldsymbol{\gamma} = \left[\gamma_1, \cdots, \gamma_{NL} \right]^T$, which characterizes the row sparsity of  $\mathbf{X}$ in CD and DD, i.e., $\gamma_n = 1$ if $\mathbf{X}(n,:)$ is nonzero and $\gamma_n \rightarrow +\infty$ otherwise. Furthermore,  let $\lambda=1\slash \sigma_n^2$ denote the noise precision, with the prior pdf following $p(\lambda) \propto 1\slash \lambda$. With the assumption that $\mathbf{X}$ is independent among codewords, channel taps, and antennas,  the joint \textit{a posteriori} pdf of $\mathbf{X}, \boldsymbol{\gamma}$, and $\lambda$ follows that 
 \begin{equation}
	\begin{aligned}
		&p(\mathbf{X}, \boldsymbol{\gamma}, \lambda | \mathbf{Y}) \propto p(\mathbf{Y} | \mathbf{X}, \lambda) p( \mathbf{X} | \boldsymbol{\gamma}) p(\lambda) p(\boldsymbol{\gamma}) \\
		&=  p(\lambda) \prod_{s=1}^{S} \prod_{m=1}^{M} {p(y_{s,m} | \mathbf{x}_m, \lambda)}  \prod_{n=1}^{NL}  p(\gamma_n) {\prod_{m=1}^{M} {p({x}_{n,m} | \gamma_n )}}, 
	\end{aligned} \label{equ-10} 
\end{equation}
where $\mathbf{x}_m$ is the $m$-th column of $\mathbf{X}, p(y_{s,m} | \mathbf{x}_m, \lambda) = \mathcal{CN}(y_{s,m}; \mathbf{g}_s^T \mathbf{x}_m, \lambda^{-1})$ and $p({x}_{n,m} | \gamma_n ) = \mathcal{CN}(x_{n,m};0,\gamma_n^{-1})$. Following the SBL principle, the prior pdf of $\gamma_n$ follows the Gamma distribution, i.e., $p(\gamma_n)= Ga(\gamma_n;\epsilon, \eta)$. To facilitate the factor graph (FG) representation, we further factorize Eq. \eqref{equ-10} by introducing auxiliary beliefs, i.e., we employ $f_{\lambda}(\lambda), f_{\gamma_n}(\gamma_n, \epsilon)$, and $ f_{x_{n,m}}(x_{n,m}, \gamma_n)$ to denote $p(\lambda), p(\gamma_n)$, and $p({x}_{n,m} | \gamma_n)$, respectively, which act as the factor nodes (FNs) in the FG and capture the prior distributions of involved variable nodes (VNs). To exploit the superiority of BP algorithm in dealing with discrete probability and linear Gaussian models, we introduce the auxiliary variable $\omega_{s,m} = \mathbf{g}_s^T \mathbf{x}_m$ and the constraint $\delta(\omega_{s,m}-\mathbf{g}_s^T \mathbf{x}_m)$, denoted by $f_{\delta_{s,m}}(w_{s,m}, \mathbf{x}_m)$. Correspondingly, $f_{y_{s,m}}(w_{s,m},\lambda) =  \mathcal{CN}(y_{s,m}; \omega_{s,m}, \lambda^{-1})$. Thus, Eq. \eqref{equ-10} can be factorized as
\begin{equation}
	\begin{aligned}
		&p(\mathbf{X}, \boldsymbol{\gamma}, \lambda | \mathbf{Y}) \propto  f_{\lambda}(\lambda )  \prod_{s=1}^{S} \prod_{m=1}^{M} f_{y_{s,m}}(w_{s,m},\lambda) \\
		&\times  f_{\delta_{s,m}}(w_{s,m}, \mathbf{x}_m)  \prod_{n=1}^{NL}  f_{\gamma_n}(\gamma_n) {\prod_{m=1}^{M} { f_{x_{n,m}}(x_{n,m}, \gamma_n)}}.
	\end{aligned} \label{equ-11}
\end{equation}

\par Instead of adopting the traditional SBL algorithm with matrix inversion operations, we leverage BP algorithm to iteratively estimate the sparse channel $\mathbf{X}$ and resort to MF algorithm to iteratively update the hyper-parameters $\mathbf{\gamma}, \lambda$, and $\epsilon$ \cite{Zhang2018TVT, Zhang2023IOT}, which is more suitable for cases where the posterior probabilities cannot be obtained directly.

\begin{figure}[htpb]
	\centerline{\includegraphics[width=0.42\textwidth]{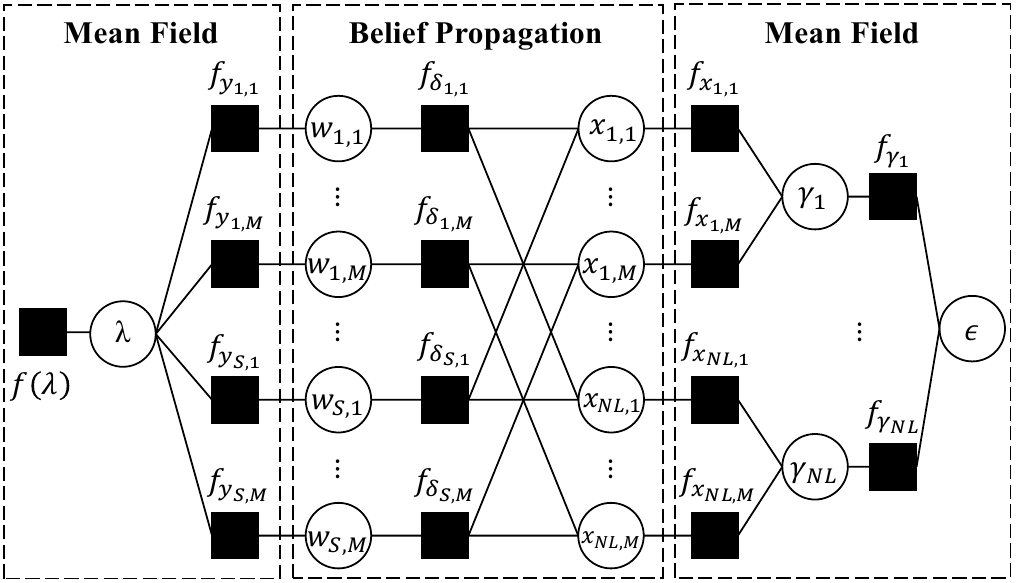}}
	\caption{FG representation of Eq. \eqref{equ-11}.}
	\label{pic-2}
\end{figure}

\par The FG is demonstrated in Fig. \ref{pic-2}, with solid squares and hollow circles denoting FNs and VNs, respectively. The BP rule is deployed for the FNs $\left\lbrace f_{\delta_{s,m}}, \forall s,m\right\rbrace$, and the MF rule is applied to the remaining FNs. The message passed from VN (FN) $a$ to FN (VN) $b$ is denoted by $I_{a \rightarrow b}(x)$, which is a function of $x$, and the belief of VN $x$ is denoted by $b(x)$. With these definitions, we elaborate on the updating rules for both forward (from right to left) and backward (from left to right) messages. For simplicity, the iteration index is omitted.

\subsubsection{Forward Message Passing} 
 
 \par Firstly, The message from $f(\lambda)$ to $\lambda$ is $I_{f_{\lambda} \rightarrow \lambda}(\lambda) \propto 1\slash \lambda$. Provided that the belief of $\lambda$, i.e., $b(\lambda)$ is known, the message from $f_{y_{s,m}}$ to $w_{s,m}$ is updated by the MF rule, which follows that $I_{f_{y_{s,m}} \rightarrow w_{s,m}}(w_{s,m}) \propto \mathcal{CN}(w_{s,m}; y_{s,m}, \hat{\lambda}^{-1})$, where $	\hat{\lambda} =  \left\langle  \lambda \right\rangle _{b(\lambda)}$. The message from $w_{s,m}$ to $f_{\delta_{s,m}}$ equals that from $f_{y_{s,m}}$ to $w_{s,m}$ , i.e., $I_{w_{s,m}\rightarrow f_{\delta_{s,m}}}(w_{s,m}) = I_{f_{y_{s,m}} \rightarrow  w_{s,m}}(w_{s,m})$. Accordingly, the message from $f_{\delta_{s,m}}$ to $x_{n,m}$ is calculated by the BP rule and follows that $I_{f_{\delta_{s,m}} \rightarrow x_{n,m}} \propto \mathcal{CN}\left( x_{n,m} ; \mu_{s \rightarrow n}^m, \nu_{s\rightarrow n}^m \right)$, where $\mu_{s \rightarrow n}^m$ and $\nu_{s\rightarrow n}^m$ are the estimated posterior mean and variance observed at FN $f_{\delta_{s,m}}$, respectively, which are given by
\begin{align}
	\mu_{s \rightarrow n}^m &= \left( {y_{s,m} - \sum\nolimits_{k\neq n} g_{s,k} \mu_{k \rightarrow s}^m }\right) \big\slash  {g_{s,n}}, \label{equ-12} \\ 
	\nu_{s\rightarrow n}^m   &=  \left( { \sum\nolimits_{k\neq n}  \left| g_{s,k}\right| ^2  \nu_{k\rightarrow s}^m  + {\hat{\lambda}}^{-1} } \right)  \big\slash {\left| g_{s,n}\right| ^2}. \label{equ-13}
\end{align}
Likewise, by combining all incoming messages, the belief of $x_{n,m}$ follows the Gaussian distribution, i.e., $b(x_{n,m}) \propto \mathcal{CN}\left( x_{n,m}; \mu_{n,m}^x, \nu_{n,m}^x \right)$, where 
\begin{align}
	\nu_{n,m}^x &= \left( \hat{\gamma}_n+ \sum\nolimits_s {1} \slash {\nu_{s\rightarrow n}^m} \right)^{-1},    \label{equ-15}\\
	\mu_{n,m}^x &= \nu_{n,m}^x  \sum\nolimits_s {\mu_{s \rightarrow n}^m} \slash {\nu_{s\rightarrow n}^m},   \label{equ-16}
\end{align}
where $\mu_{n,m}^x$ and $\nu_{n,m}^x$ are the posterior estimation and the variation of $\mathbf{X}(m,n)$, respectively, and the channel precision $\hat{\gamma}_n$ is updated in Eq. \eqref{equ-18}. With the combination of all related messages, the belief of $\gamma_n$ follows that 
\begin{equation}
	\begin{aligned}
		& b({\gamma_n}) \propto I_{f_{\gamma_n} \rightarrow \gamma_n}(\gamma_n) \prod_m I_{f_{x_{n,m}} \rightarrow \gamma_n}(\gamma_n )\\
		&\propto \gamma_n^{\hat{\epsilon} + M -1} \exp \Big\{- \gamma_n \Big[ \eta + \sum_m \big( \left|  \mu_{n,m}^x\right|^2 + \nu_{n,m}^x \big) \Big] \Big\},
	\end{aligned} \label{equ-17}
\end{equation}
where $I_{f_{\gamma_n} \rightarrow \gamma_n}(\gamma_n)  \propto Ga (\gamma_n; \hat{\epsilon}, \eta)$ is the prior pdf of $\gamma_n$. Consequently, $\gamma_n$ is estimated by
\begin{equation}
	\hat{\gamma}_n =  {\left( \hat{\epsilon}+M\right) }\Big\slash {\Big( \eta +  \sum_m \left( \left|  \mu_{n,m}^2\right|^x + \nu_{n,m}^x \right)\Big)}.   \label{equ-18}
\end{equation}
Since the closed form for updating the hyper-parameter $\hat{\epsilon}$  is challenging to derive, an empirical yet effective solution for $\hat{\epsilon}$ is given by \cite{epsilon}
\begin{equation}
	\hat{\epsilon}  = \frac{1}{2} \sqrt{\ln \left( \frac{1}{NL}\sum\nolimits_{n} \hat{\gamma}_n \right)  -  \frac{1}{NL} \sum\nolimits_n \ln \hat{\gamma}_n }.  \label{equ-20}
\end{equation}

\subsubsection{Backward Message Passing}
\par As mentioned above, the backward message from $f_{\gamma_n}$ to $\gamma_n$ is the prior pdf of $\gamma_n$ with the hyper-parameter $\hat{\epsilon}$ updated according to Eq. \eqref{equ-20}. And the message from $f_{x_{n,m}}$ to $x_{n,m}$ follows the contribution that $I_{f_{x_{n,m}} \rightarrow x_{n,m}}(x_{n,m})  \propto \mathcal{CN}(x_{n,m}; 0, \hat{\gamma}_n^{-1})$.  Likewise, the message from $x_{n,m}$ to $f_{\delta_{s,m}}$  following the Gaussian distribution, i.e., $I_{x_{n,m} \rightarrow f_{\delta_{s,m}}}(x_{n,m})  \propto \mathcal{CN}\left( x_{n,m} ; \mu_{n \rightarrow s}^m, \nu_{n \rightarrow s}^m \right)$, where $\mu_{n \rightarrow s}^m$ and $\nu_{n \rightarrow s}^m$ are the estimated posterior mean and variance observed at VN $x_{n,m}$, respectively, which reads
\begin{align}
	\mu_{n \rightarrow s}^m &= \nu_{n \rightarrow s}^m \left({\mu_{n,m}^x}\slash {\nu_{n,m}^x}  - {\mu_{s \rightarrow n}^m}\slash {\nu_{s\rightarrow n}^m}\right),  \label{equ-21} \\
	\nu_{n \rightarrow s}^m &= \left({1} \slash {\nu_{n,m}^x}  - {1} \slash {\nu_{s\rightarrow n}^m} \right)^{-1}.  \label{equ-22}	
\end{align}
Similar to message  $I_{x_{n,m} \rightarrow f_{\delta_{s,m}}}$, the message  from $f_{\delta_{s,m}}$ to $w_{s,m}$ follows the Gaussian distribution, with the mean and variance updated by
\begin{align}
	\mu_{\delta \rightarrow w}^{s,m} &= \sum\nolimits_n g_{s,n} \mu_{n \rightarrow s}^m, \label{equ-23}\\
	\nu_{\delta \rightarrow w}^{s,m}  &= \sum\nolimits_{n}  \left| g_{s,n}\right| ^2  \nu_{n\rightarrow s}^m.  \label{equ-24}
\end{align}
Correspondingly, the belief of $w_{s,m}$ is derived as $b(w_{s,m}) \propto \mathcal{CN}(w_{s,m}; \mu_{w_{s,m}}, \nu_{w_{s,m}})$, where
\begin{align}
	\nu_{{s,m}}^w &= \left(\hat{\lambda} + {1}\slash {\nu_{\delta \rightarrow w}^{s,m}} \right)^{-1}, \label{equ-26}\\
	\mu_{{s,m}}^w  &= \nu_{{s,m}}^w \left(y_{s,m}\hat{\lambda} +  {\mu_{\delta \rightarrow w}^{s,m}}\slash {\nu_{\delta \rightarrow w}^{s,m}}\right). \label{equ-27}
\end{align}
In this way, the belief of $\lambda$ can be obtained according to the MF rule, which is given by
\begin{equation}
	 b(\lambda)\propto \lambda^{SM - 1} \exp \Big\{ \! -\lambda \sum_{s,m} \left(\left| y_{s,m} \!-\! \mu_{{s,m}}^w\right|^2 \!+\! \nu_{{s,m}}^w  \right) \! \Big\}.  \label{equ-28}
\end{equation}
Finally, based on the posterior expectation, $\hat{\lambda}$ is updated by
\begin{equation}
	\hat{\lambda} = \left\langle  \lambda\right\rangle _{b(\lambda)} = SM \slash \sum_{s,m} \left(\left| y_{s,m} - \mu_{{s,m}}^w\right|^2 + \nu_{{s,m}}^w  \right). \label{equ-29}
\end{equation}
The above derivations are summarized as the JADCE-MP-SBL algorithm in Alg. \ref{alg-1}.  Messages are exchanged iteratively until the maximum number of iterations is reached. Besides, we leverage the normalized mean squared error (NMSE) (see line \ref{line-14}) as another stopping criterion for a certain tolerance $\tau$. We make a hard decision to detect the active rows with an appropriate threshold $v$, i.e., $\widehat{\mathcal{I}}^t =  \lbrace I:  {|| \widehat{\mathbf{X}}^t(I,:)||_2^2 > v}, I \in [NL] \rbrace$, where $|\widehat{\mathcal{I}}^t| \leq \sum\nolimits_{k} L_{k}$ due to possible codeword collisions.

 \begin{algorithm} [htpb]
	\caption{JADCE-MP-SBL Algorithm}
	\label{alg-1}  
	\begin{algorithmic}[1]	
		\STATE {{\bf Input}: $\mathbf{G}, \mathbf{Y}^t, \forall t \in [T_p]$}\\
		\STATE{{\bf Initial}: $\forall n: \hat{\gamma}_n = 1, \hat{\epsilon} = 10^{-3}, \hat{\lambda} = 1, \eta = 10^{-4}$\\
			\quad \quad \quad \;$\forall n,m,s: \nu_{n \rightarrow s}^m = 1, \mu_{n \rightarrow s}^m = 0$}
		\STATE {{\bf for} $i=1: N_{iter}$ {\bf do}}
		\STATE {\quad \% {\textbf{Forward Message Passing}}}
		\STATE {\quad  Update $\mu_{s \rightarrow n}^m$ and $\nu_{s\rightarrow n}^m$ with \eqref{equ-12} and \eqref{equ-13}, $\forall n,m,s$}
		\STATE {\quad  Update  $\nu_{n,m}^x$ and $\mu_{n,m}^x$ with \eqref{equ-15} and \eqref{equ-16}, $\forall n,m$}
		\STATE {\quad \% {\textbf{Backward Message Passing}}}
		\STATE {\quad  Update $\mu_{n \rightarrow s}^m$ and $\nu_{n \rightarrow s}^m$ with \eqref{equ-21} and \eqref{equ-22}, $\forall n,m,s$}
		\STATE {\quad  Update $\mu_{\delta \rightarrow w}^{s,m}$ and $\nu_{\delta \rightarrow w}^{s,m}$ with \eqref{equ-23} and \eqref{equ-24}, $\forall m,s$}
		\STATE {\quad  Update $\nu_{{s,m}}^w$ and $\mu_{{s,m}}^w$ with \eqref{equ-26} and \eqref{equ-27}, $\forall m,s$ }
		\STATE {\quad \% {\textbf{Parameter Updating}}}
		\STATE {\quad Update $\hat{\gamma}_n$ with \eqref{equ-18}, $\hat{\epsilon}$ with \eqref{equ-20}, $\hat{\lambda}$ with \eqref{equ-29}}
		\STATE {\quad  Estimated channel $\hat{x}_{n,m}^t(i)  = \mu_{n,m}^x, \forall n,m$}
		\STATE{\quad {\bf if} $\| \widehat{\mathbf{X}}^t(i+1) - \widehat{\mathbf{X}}^t(i) \|_F^2 < \tau \| \widehat{\mathbf{X}}^t(i) \|_F^2$, {\bf stop}} \label{line-14}
		\STATE {{\bf end for}}  
		\STATE {{\bf Output}: {Estimated channel  $\widehat{\mathbf{X}}^t, \forall t\in [T_p]$}}
	\end{algorithmic}  
\end{algorithm}

\par It is noteworthy that the JADCE-MP-SBL algorithm can iteratively learn the activity probability, channel, and noise precision without requiring prior knowledge. In contrast to the AMP algorithm \cite{Liu2018tsp}, which necessitates such prior knowledge to ensure performance, JADCE-MP-SBL can achieve comparable performance under more relaxed conditions. Moreover, as illustrated in Eq. \eqref{equ-38}, the collision threshold can be determined based on the estimated precision of the channel and noise, rather than being selected empirically, which is crucial in algorithm design. Furthermore, JADCE-MP-SBL performs CE by iteratively calculating forward and backward messages, effectively circumventing the need for matrix inversion operations. These calculations can be decomposed into local tasks and executed in parallel, thereby significantly reducing both computational and time complexities.
 
\subsection{GB-CR$^2$ Algorithm} \label{sec-IV-B}

\par
\begin{figure}[htpb]
	\centerline{\includegraphics[width=0.38\textwidth]{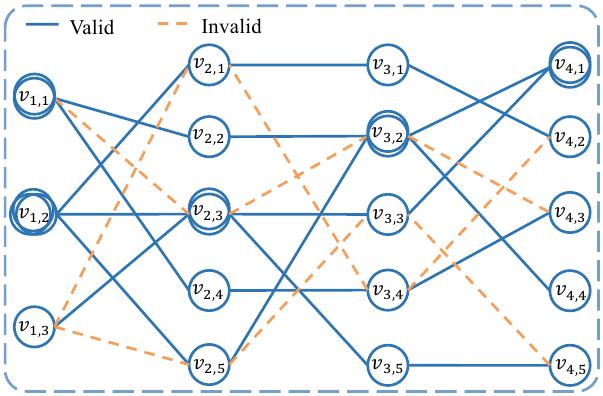}}
	\caption{An example of the proposed graph with $K_a=6$ and four stages, with both valid (solid blue) and invalid (dotted orange) edges. The overlapped circles correspond to the collision cases.}
	\label{pic-3}
\end{figure}

 For tractability, we resort to the FD channel $\widehat{\mathbf{X}}_n^t\in \mathbb{C}^{S\times M}, \forall n\in [N], t \in [T_p]$ for subsequent operations, which is reconstructed as below
\begin{equation}
	\widehat{\mathbf{X}}_n^t (s,m) = \sum\nolimits_I \mathbb{I}_{l = n} \widehat{\mathbf{X}}^t(I,m) e^{-j2\pi (n_s-1) \lceil I / N\rceil  / N_c}, \label{equ-30}
\end{equation}
where  $I \in\widehat{\mathcal{I}}^t, n \in [N]$, $l \triangleq \text{mod} [(I-1)/N]+1$ and $\lceil \cdot \rceil  $ denotes the module and ceiling operations, respectively. $\mathbb{I}$ is the indicator function, i.e., $\mathbb{I}_{l = n}= 1$ if $l=n$ and $0$ otherwise. Note that $\widehat{\mathbf{X}}_n^t$ cannot be applied for data decoding directly, since it is coupled with the CFO-caused phase rotation $p_k^t$ and superimposed by the channel taps of all collided users. The solution to these two problems forms the content of the proposed GB-CR$^2$ algorithm, which iteratively separates the superimposed channels across multi-stage preambles with a moderate codebook dimension. The algorithm design revolves around two specific tasks: 1) cross-segment splicing of user data and 2) channel reconstruction along with collision resolution and CFO compensation. The illustration begins with the definition of associated items. 
\begin{itemize}
	\item \textit{Node:} The user's data at each stage. Nodes are associated with channels and connected across stages through edges. Overlapped nodes indicate the collided codewords.
	\item \textit{Edge:}  Signifies \textit{possible} connections between nodes, initialized in the tree decoding process by the parity checks.
	\item \textit{Weight:} Reflects the credibility of an edge. A larger weight indicates less reliability of the edge.
	\item \textit{Path:} Encompasses the connected nodes and edges across all stages. 
\end{itemize}
We further introduce variables $\{ N_t, \mathbf{v}_t \}_{t=1}^{T_p} $ and the set $\mathcal{P}$ to characterize the graph, where $N_t$ denotes the number of nodes in the $t$-th stage, and $\mathbf{v}_t = [v_{t,1}, \cdots, v_{t,N_t}]^T \in \mathbb{Z}^{N_t\times1}$ is the vector of users' decimal messages (nodes). Note that $v_{t,n_t} = \text{mod} [(I-1)/N]+1$, where $n_t \in [N_t], I \in \widehat{\mathcal{I}}^t$. The set $\mathcal{P}$ collects the tree-decoded paths spliced based on the parity checks \cite{Amal2020TIT}, i.e., a path $\{v_{1,n_1}, \cdots, v_{T_p,n_{T_p}} \} \in \mathcal{P}$ if these nodes are stitched together after tree decoding. Specifically, we provide an example of a graph with four preamble stages in Fig. \ref*{pic-3} \footnote{Note that  there should be $C_4^2=6$ edges on each path, while for the clarity of illustration, we only draw the edges between adjacent nodes.}, where $\{v_{1,1}, v_{2,2}, v_{3,2}, v_{4,1} \} \in \mathcal{P}$ since these nodes are connected through edges, and $\{v_{1,1}, v_{2,1}, v_{3,1}, v_{4,1}\} \notin \mathcal{P}$ since no path goes through these nodes.
Note that invalid paths exist due to the limited error correction capabilities of tree coding, leading to EPs in the prior arts \cite{Amal2020TIT}. In this paper, the tree code is a coarse and auxiliary method, striking a balance between decoding complexity and unguaranteed accuracy. The validity of paths is further scrutinized by the GB-CR$^2$ algorithm. Intuitively, the channel MSE between the nodes on a valid path is small, since the channel for a user remains constant across multiple symbols. Following this principle, we define the weight of an edge as the MSE between the FD channels on the nodes it connects. Note that the phase rotation needs to be eliminated when calculating the MSE since it varies on symbols. For tractability, we assume that the CFO is within $\mathbf{q} = [\epsilon^{(1)}, \epsilon^{(2)} \cdots, \epsilon^{(Q)}]$, uniformly sampled from the range $[-\epsilon_{\max}, \epsilon_{\max}]$ with the quantization level $Q$. Thus, the weight between nodes $v_{i,n_i}$ and $v_{j,n_j}$ with the $n$-th quantized CFO $\epsilon^{(n)}$ is given by
\begin{equation}
	w_{j,n_j}^{i,n_i}(\epsilon^{(n)})  = \left\| \frac{1}{{q}^i(\epsilon^{(n)})}\widehat{\mathbf{X}}^i_{v_{i,n_i}}  - \frac{1}{{q}^j(\epsilon^{(n)})}\widehat{\mathbf{X}}^j_{v_{j,n_j}} \right\|_F^2, \label{equ-31}
\end{equation}
where $q^i(\epsilon) = \omega^{(L+N_c)i- {(N_c+1)}\slash{2}} $ with $\omega = e^{j2\pi \epsilon/N_c}$. Consequently, the total weights $\Omega$ on $p\in \mathcal{P}$ and $\epsilon^{(n)}$ is given by
\begin{equation}
	\Omega(p, \epsilon^{(n)}) = \sum\nolimits_{1\leq i <j \leq T_p}w_{j,n_j}^{i,n_i}(\epsilon^{(n)}),
\end{equation}
where $p=\{v_{1,n_1},\cdots, v_{T_p, n_{n_{T_p}}}\}$. Generally, we can restore a path and estimate the CFO by the minimum weight path (MWP) search, i.e.,  
\begin{equation}
	\left[ \hat{p}_k, \hat{\epsilon}_k \right] = \arg \min_{p\in \mathcal{P}, \epsilon \in \mathbf{q}} \Omega (p,\epsilon). \label{equ-33}
\end{equation}
Note that $\hat{p}_k$ with the highest reliability is not necessarily a valid path. We further validate $\hat{p}_k$ through Eq. \eqref{equ-34} as follows.

\begin{Proposition} \label{prop-1}
	Given a path $\hat{p}_k$ restored from Eq. \eqref{equ-33}, for its arbitrary nodes ${v}_{i,n_i}$ and ${v}_{j,n_j}$, we consider $\hat{p}_k$ is valid if 
	\begin{equation}
			\big\| \widehat{\underline{\mathbf{X}}}_i- \widehat{\underline{\mathbf{X}}}_j \big\|_F^2 <  \max \left\lbrace \big\| \widehat{\underline{\mathbf{X}}}_i\big\|_F^2, \big\|\widehat{\underline{\mathbf{X}}}_j  \big\|_F^2 \right\rbrace, \label{equ-34}
	\end{equation}
where $\widehat{\underline{\mathbf{X}}}_i  \triangleq \widehat{\mathbf{X}}^i_{{v}_{i,n_i}}\slash q^i(\hat{\epsilon}_k)$ and $\widehat{\underline{\mathbf{X}}}_j   \triangleq \widehat{\mathbf{X}}^j_{{v}_{j,n_j}}\slash q^j(\hat{\epsilon}_k)$. Based on this criterion, we can bound the probability of the event that a valid path is falsely detected to be invalid as

\begin{equation}
	  P \left\lbrace \big\| \widehat{\underline{\mathbf{X}}}_i- \widehat{\underline{\mathbf{X}}}_j \big\|_F^2 > max \left\lbrace \big\| \widehat{\underline{\mathbf{X}}}_i\big\|_F^2, \big\|\widehat{\underline{\mathbf{X}}}_j  \big\|_F^2 \right\rbrace  \right\rbrace < \delta,  \label{equ-35}
\end{equation}
where 
\begin{equation}
	\frac{\sigma^2}{\max\{\sigma_i^2, \sigma_j^2\}} < \frac{ \Gamma_{2SM}^{-1} (\delta\slash 2)}{\Gamma_{2SM}^{-1}(1-\delta \slash 2)},  \label{equ-36}
\end{equation}
and $\sigma_i^2,\sigma_j^2$, and $\sigma^2$ are the variance of $\widehat{\underline{\mathbf{X}}}_i, \widehat{\underline{\mathbf{X}}}_j$, and $\widehat{\underline{\mathbf{X}}}_i-\widehat{\underline{\mathbf{X}}}_j$, respectively. $ \Gamma_k(\cdot)$ is the cumulative distribution function (CDF) of chi-squared distribution with $k$ degrees of freedom $\chi_{k}^2$, and $\Gamma_k^{-1}$ is its inverse.
\begin{IEEEproof}
	Please see Appendix \ref{app-1}.
\end{IEEEproof}
\end{Proposition}

\begin{Remark}
Proposition \ref{prop-1} introduces an upper bound $\delta$ of the probability that a valid path violates the rule in Eq. \eqref{equ-34}. For a valid path, $\sigma^2 < \max\{\sigma_i^2,\sigma_j^2\}$ always holds. Note that the stronger condition $\sigma^2 \ll \max\{\sigma_i^2,\sigma_j^2\}$ holds for non-collided nodes in most cases, since the channel can be eliminated and only noise remains. In such cases, the upper bound $\delta$ to satisfy Eq. \eqref{equ-36} can be further reduced, particularly for large $S$ and $M$, which underscores the effectiveness of the validity criterion for paths in Eq. \eqref{equ-34}. 
\end{Remark}
% 公式47的概率 \delta是一个概率上界，而公式48表明，当p是有效路径时，不等式左边会变小，而右边是关于\delta的单调递增函数，因此满足公式48条件的delta可以变小，因此可以降低概率上界，从而表明p有效时，不满足判断准则的概率很小，从而表明判断标准是有效的。

\par Furthermore, the non-collided node $v_{i,n_i}$ of a valid path $\hat{p}$ should satisfy that
\begin{equation}
	\big\| \widehat{\underline{\mathbf{X}}}_i- \widehat{\underline{\mathbf{X}}}_j \big\|_F^2  \leq \gamma_{c},  \label{equ-37}
\end{equation}
where $v_{j,n_j}$ is the node of $\hat{p}$ with the lowest channel energy, and $\gamma_{c}$ is obtained by Neyman Pearson (NP) hypothesis testing. Note that the node $v_{j,n_j}$ serves as a reference point for evaluating the channel's energy. Compared to collision detection methods that rely solely on the channel's own energy, Eq. \eqref{equ-37} introduces a novel criterion for detecting collisions based on the channel's relative energy across multiple slots, which is more accurate and reasonable in the presence of large-scale fading. The threshold $\gamma_{c}$ in Eq. \eqref{equ-37} is given by
\begin{equation}
	\gamma_{c}=  \hat{\sigma}_e^2 \Gamma_{2SM}^{-1}(1-\zeta),   \label{equ-38}
\end{equation}
where $\zeta$ is the test level and also the bound on the false-alarm probability, and $\hat{\sigma}_e^2$ is the Bayesian Cramér-Rao bound (BCRB) of the MSE later given in Eq. \eqref{equ-51}, with $\boldsymbol{\Sigma}_{\mathbf{n}}=\hat{\lambda}^{-1}\mathbf{I}_{S}$ and $\boldsymbol{\Sigma}_{\mathbf{x}}= \text{diag}\{\hat{\gamma}_n^{-1}, n\in \hat{\mathcal{I}}\}$. For the proof of Eq. \eqref{equ-38}, please see Appendix \ref{app-2}. After validating the minimum-weight path and determining the non-collided nodes, we can readily retrieve the channel as given by
\begin{equation}
	\widehat{\mathbf{H}}_k = 1\slash \left|\mathcal{V}_p \right|  \sum\nolimits_{\mathbf{v}_i(n_i) \in \mathcal{V}_p}{\widehat{\underline{\mathbf{X}}}_i^T},  \label{equ-39}
\end{equation}
where $\mathcal{V}_p$ denotes the set of non-collided nodes of path $p$ and $\widehat{\mathbf{H}}_k$ is then eliminated in the collided nodes of $p$. Once a path is picked up, the corresponding edges will be deleted in the graph. Besides, all other paths connected to the non-collided nodes in the selected path are also deleted. Thus, at least one path is removed from the graph in each iteration. The iteration terminates when all paths on the graph are deleted or no path satisfies the validity criterion in Eq. \eqref{equ-34}. Since the initial number of paths is finite, the algorithm is guaranteed to converge, with the number of retrieved paths corresponding to the estimated number $\hat{K}_a$ of active users. The convergence is a byproduct of the algorithm adopted to estimate $\hat{K}_a$, which is not known in advance. The overall algorithm is summarized in Alg. \ref{alg-2}.

\par
	\begin{Remark}
		The GB-CR$^2$ algorithm exhibits a noteworthy ability to decouple the TO and CFO estimation embedded in the high-dimensional codebook into multiple slots, thereby reducing the codebook dimension. In contrast to the codebook-enlarging approaches \cite{sun2022TWC, Amal2019ICASSP,Bian2023GC}, which exhibit quadratic complexity w.r.t. the quantization level, GB-CR$^2$ achieves superior performance with only linear complexity. Additionally, by comparing the relative energy of channels among multiple slots, GB-CR$^2$ establishes a ``relative" threshold for detecting collisions (cf. Eq. \eqref{equ-37}) based on the distribution, which is more efficient than exploiting the empirical thresholds based on the absolute channel energy \cite{li2022JSAC, Guo2024IOTJ}.
	\end{Remark}

\begin{figure*}[t]
	\centerline{\includegraphics[width=0.98\textwidth]{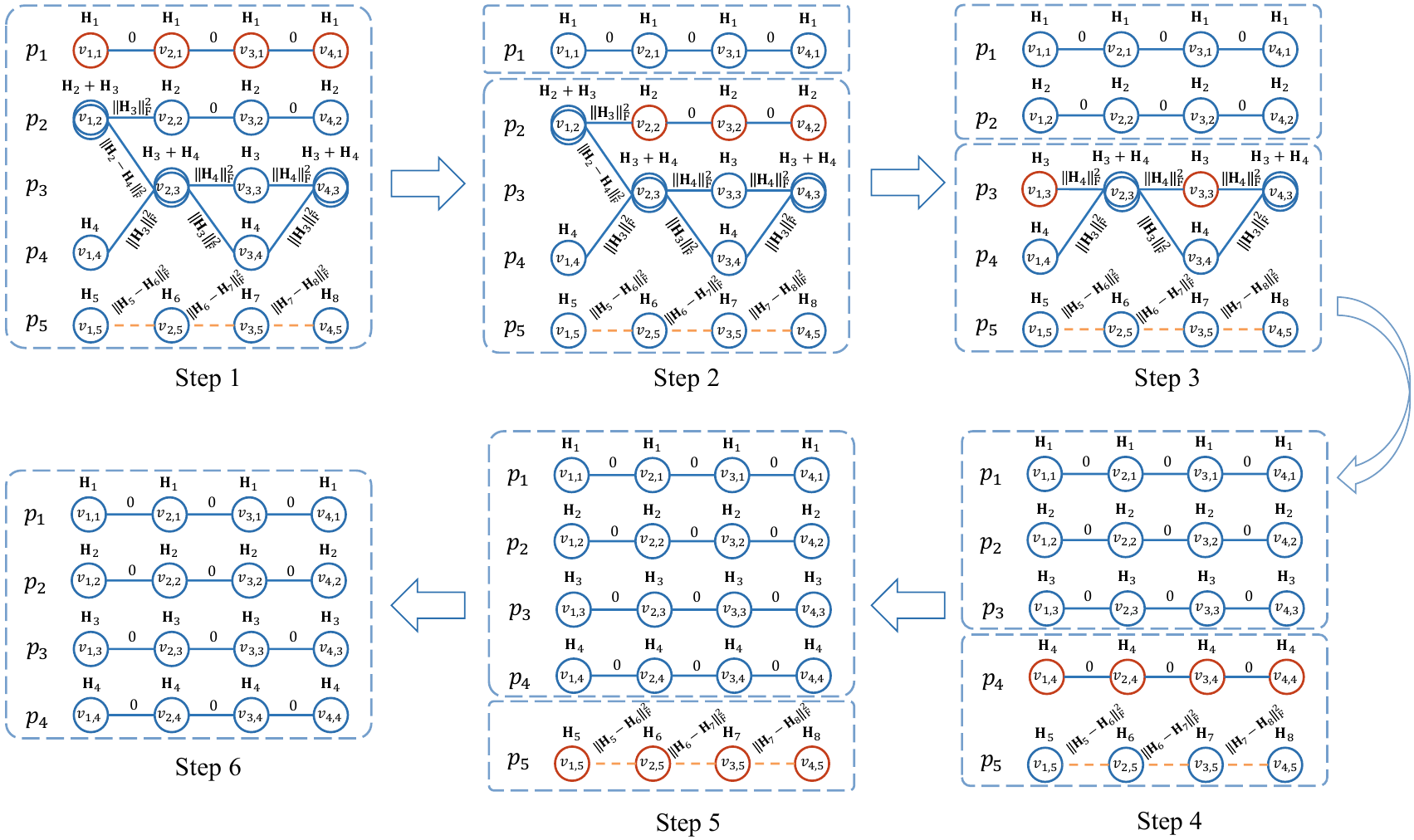}}
	\caption{ An example of the implementation process of GB-CR$^2$ algorithm with four stages and five paths.}
	\label{pic-5}
	\vspace{-0.2cm}
\end{figure*}

 \begin{algorithm} [htpb]
%	\setstretch{1.15}
	\caption{The Proposed GB-CR$^2$ Algorithm}
	\label{alg-2}  
	\begin{algorithmic}[1]	
		\STATE {{\bf Input}: $\mathcal{P}$, $\mathbf{d}$, $\mathbf{q}$, $\mathbf{v}_i, \widehat{\mathbf{X}}^t_n, \forall n\in [1:N_t], t\in [1:T_p]$}\\
		\STATE{{\bf Initial}: $ \widehat{\mathcal{P}}= \emptyset$, $\widehat{K}_a=0$}
		\REPEAT
		\STATE {{ \% \textbf{Minimum Weight Path Searching}}}\\
		\STATE {\% {For FSF Channel}}
		\STATE {\quad Update: $\left(\hat{p}, \hat{\epsilon}_k\right)$ via Eq. \eqref{equ-33}, $\mathcal{P} \leftarrow \mathcal{P} - \hat{p}$}\\
		\STATE {\% {For Flat Fading Channel}}
		\STATE {\quad Update: $\left(\hat{p},\hat{\tau}_k, \hat{\epsilon}_k\right)$ via Eq. \eqref{equ-43}, $\mathcal{P} \leftarrow \mathcal{P} - \hat{p}$}\\
		\STATE{ {For valid path $\hat{p}$ fulfilling Eq. \eqref{equ-34}}}\\
		\STATE {\quad Update: $\widehat{\mathbf{H}}_k$ via Eq. \eqref{equ-39}, $\widehat{K}_a \leftarrow \widehat{K}_a+1, \widehat{\mathcal{P}} \leftarrow \widehat{\mathcal{P}} \cup \hat{p} $}\\
		\STATE {\quad \% \textbf{Successive Interference Cancellation}}\\
		\STATE{ \quad {For collided node $v_{i,n_i}$ NOT fulfilling Eq. \eqref{equ-37}}}\\
		\STATE{\quad \quad  $\widehat{\mathbf{X}}^i_{v_{i,n_i}} \leftarrow \widehat{\mathbf{X}}^i_{v_{i,n_i}} -  q^i(\hat{\epsilon}_k)\widehat{\mathbf{H}}_k$   \% { FSF  }}\\
		\STATE{\quad \quad $\widehat{\mathbf{x}}^i_{v_{i,n_i}} \leftarrow\widehat{\mathbf{x}}^i_{v_{i,n_i}} - q^i_{v_{i,n_i}}(\hat{\tau}_k, \hat{\epsilon}_k) \widehat{\mathbf{h}}_k$  \% Flat Fading}\\ \label{line-14-2}
		\UNTIL {$\mathcal{P}=\emptyset$}
		\STATE {\% \textbf{Re-CE For Flat Fading Channel with Known Offsets}}\\
		\STATE {\quad Update: Optimal $(\hat{\tau}_k, \hat{\epsilon}_k)$ via Eq. \eqref{equ-49}}\\
		\STATE {\quad Update: $\widehat{\mathbf{H}}_k$ via Eq. \eqref{equ-39} and channel SIC via line \ref{line-14-2}}\\
		\STATE {{\bf Output}: {$ \widehat{\mathcal{P}}$, $\widehat{\mathbf{H}}_k$, $\hat{\tau}_k$, $\hat{\epsilon}_k, \forall k\in [1:\widehat{K}_a]$}}
	\end{algorithmic}  
\end{algorithm}

\par We provide an illustrative example to demonstrate the working procedure of the GB-CR$^2$ algorithm. As shown in Fig. \ref{pic-5}, there are five paths, i.e., $\mathcal{P}=\{p_1,\cdots, p_5\}$, where non-collided nodes on each path, denoted by red circles, are utilized for CE. In Step $1$, the weights of all edges are initialized by calculating the channel MSE. Consequently, $p_1$ is chosen since it has the minimum weight after the MWP search. All nodes on $p_1$ are non-collided and thus utilized for channel reconstruction according to Eq. \eqref{equ-39}. The algorithm then proceeds to Step $2$, dealing with the four remaining paths in the current graph. Similarly, after the MWP search, $p_2$ is selected, and its non-collided nodes $\{v_{2,2}, v_{3,2}, v_{4,2}\}$ are employed for CE, where the estimated channel is then eliminated in the collided node $v_{1,2}$. The algorithm iteratively proceeds to Step $5$, where only $p_5$ exists in the current graph. Following the criterion in Eq. \eqref{equ-34}, $p_5$ is detected as invalid and thus discarded. Finally, the algorithm outputs four valid paths, along with reconstructing the channels as well as estimated CFOs.

\section{Application To Flat Fading Channels} \label{sec-4}

\par  Note that for the subcarrier spacing $\Delta f = 15$ kHz, when only a small fraction of subcarriers are occupied in an OFDM system, it can be  regarded as a narrowband system and the channel can be assumed as flat among subcarriers  \cite{sun2022TWC, yury2019ACSSC, Decu2022GC}. To further shed light on the effectiveness and versatility of the proposed algorithm, we apply GB-CR$^2$ to the flat fading scenario, where a significant challenge lies in that the estimation of both TO and CFO results in the parameter dimension increase and  accuracy degradation. To this end, we innovatively leverage the geometric characteristics of the signal constellation to correct the estimated offsets and enhance CE results. Specifically, the FD channel matrix $\mathbf{H}_k\in\mathbb{C}^{S\times M}$ in Eq. \eqref{equ-9} reduces to the vector $\mathbf{h}_k\in\mathbb{C}^{M\times 1}$, and the received signal is given by
\begin{equation}
	\mathbf{Y}^t = \sum_{k=1}^{K_a} \text{diag}\{\mathbf{p}_k^t\} \mathbf{A} \mathbf{c}_k^t \mathbf{h}_k^T + \mathbf{Z}.
\end{equation}

\par
\begin{figure}[htpb]
	\centerline{\includegraphics[width=0.48\textwidth]{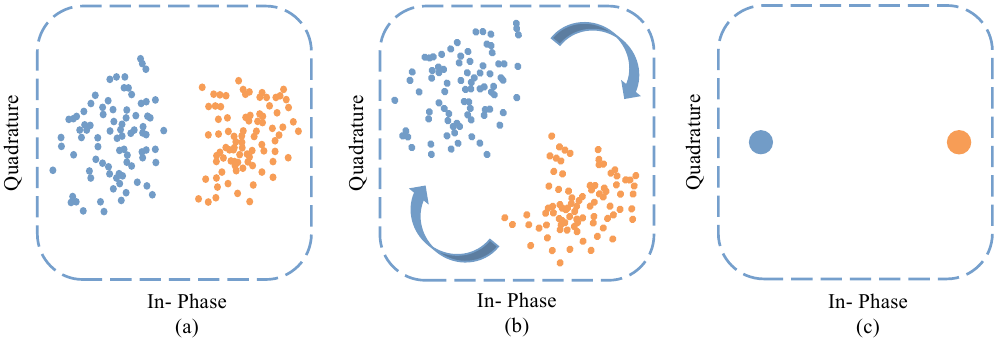}}
	\caption{Three cases of the constellation of BPSK modulation. (a) Perfect rotation compensation. (b) Imperfect compensation with residual phase rotation and $\left\| \rho_a \right\|> \left\| \rho_b\right\|$. (c) The constellation of $\underline{\mathbf{s}}_{k}^c$.}
	\label{pic-6}
\end{figure}

 Since the channel responses remain flat among nearby subcarriers, only a small number of pilots are required to reconstruct the FD channel. Therefore, to reduce multi-user interference and computational complexity, we utilize the extremely sparse orthogonal pilots (ESOPs), i.e., the identity matrix, as the codebook. In comparison to the JADCE-MP-SBL algorithm, which requires multiple iterations to converge, the linear minimum mean square error (LMMSE) estimator demonstrates superior performance and lower complexity with ESOPs. Hence, we utilize the LMMSE estimator for CE in the narrowband scenario, with the estimated channel given by
\begin{equation}
	\widehat{\mathbf{X}}^t = \mathbf{A}^H\left(\mathbf{A}\mathbf{A}^H + \sigma_n^2 \mathbf{I}_S \right)^{-1} \mathbf{Y}^t \triangleq {\mathbf{Y}^t}\slash{(\sigma_n^2+1)},
\end{equation}
where $\mathbf{A} = \mathbf{I}_{N}$ with $N=L_p$, and $	\widehat{\mathbf{X}}^t = [(\hat{\mathbf{x}}_1^t)^T, \cdots, (\hat{\mathbf{x}}_N^t)^T]^T \in \mathbb{C}^{N\times M}$ with $\hat{\mathbf{x}}_n^t$ given by
\begin{equation}
	\hat{\mathbf{x}}_n^t \triangleq \sum\nolimits_{k\in\mathcal{K}_n}{\mathbf{p}^t_{k}(n)\mathbf{h}_k} + \mathbf{z},
\end{equation}
where $\mathcal{K}_n = \{ k | c_k^t(n) = 1, k\in [K_a]\}$, $\mathbf{z} \in \mathbb{C}^{N\times 1}$ is the estimation error. Similarly, $\widehat{\mathbf{X}}^t$ is a mixture of the user channels when $|\mathcal{K}_n|>1$. Besides, it is coupled with the phase rotations of both TO and CFO, adding one more dimension to the parameter estimation. By quantizing both TO and CFO, the GB-CR$^2$ algorithm works  as follows
\begin{equation}
		\left[ \hat{p}_k, \hat{\tau}_k, \hat{\epsilon}_k \right] = \arg \min_{p\in \mathcal{P}, \tau\in \mathbf{d}, \epsilon \in \mathbf{q}} \Omega (p,\tau, \epsilon), \label{equ-43}
\end{equation}
where $\mathbf{d} = [1,2,\cdots, D]^T$ is the integer-sampled TO, which is within one CP length, and $\mathbf{q}$ is the quantized CFO as above, and the weight for the $m$-th and $n$-th quantized TO and CFO is defined as follows
\begin{align}
	w_{j,n_j}^{i,n_i}(\tau^{(m)}, \epsilon^{(n)}) &= \left\|  \frac{\hat{\mathbf{x}}_{v_{i,n_i}}^i}{q^i_{v_{i,n_i}}(m,n) }  - \frac{\hat{\mathbf{x}}_{v_{j,n_j}}^j}{q^j_{v_{j,n_j}}(m,n)}\right\|_2^2, \\
	\Omega(p, \tau^{(m)}, \epsilon^{(n)}) &= \sum\nolimits_{1\leq i <j \leq T_p}w_{j,n_j}^{i,n_i}(\tau^{(m)}, \epsilon^{(n)}).
\end{align}
where $q^i_{v_{i,n_i}}(m,n)$ is the abbreviation of $q^i_{v_{i,n_i}}(\tau^{(m)},\epsilon^{(n)})$, which equals to $\mathbf{p}_k^i(v_{i,n_i})$ with $\psi = e^{i2\pi {\tau}^{(m)}\slash N_c}$ and $\omega = e^{j2\pi \epsilon^{(n)}\slash N_c}$. Similarly, following the GB-CR$^2$ algorithm, we can readily estimate the offsets and channels. However, due to the quantization errors and the limited observations of orthogonal pilots, the offset estimation of Eq. \eqref{equ-43} is not always optimal. Moreover, our numerical results illustrate that the TO estimation errors significantly degrade the system performance. Note that the phase rotations also occur in the coding part, spanning more OFDM symbols and providing more observations. Thus, the data in the coding part can be utilized to modify the offset estimation obtained by the GB-CR$^2$ algorithm. The received signal in the coding part $\mathbf{Y}_c\in\mathbb{C}^{L_c\times M}$ is given by
\begin{equation}
		\mathbf{Y}_c = \sum\nolimits_{k=1}^{K_{a}}{\mathbf{P}_{k}  f\left(\underline{\mathbf{s}}_k^c \right)  \mathbf{h}_k^T} + \mathbf{Z},
\end{equation}
where  $\mathbf{P}_{k} = \text{diag}\{[( \mathbf{p}_{k}^{(T_p+1)}) ^T, \cdots, (\mathbf{p}_{k}^{(T_p+T_d)})^T]^T\}$ and $f(\mathbf{\underline{\mathbf{s}}}_k^c)$ refers to the encoding process of the coding part, described in Section \ref{sec-3-1}, on $\underline{\mathbf{s}}_k^c \in \mathbb{C}^{\underline{L}_c\times 1}$, the modulated symbol of user $k$. After the LMMSE estimation, the estimated symbol $\widehat{\mathbf{S}} \in \mathbb{C}^{\hat{K}_a\times L_c}$ is given by
\begin{equation}
	\widehat{\mathbf{S}}  = \widehat{\mathbf{H}}^* ( \widehat{\mathbf{H}}^T \widehat{\mathbf{H}}^* + \sigma_n^2 \mathbf{I}_M )^{-1} \mathbf{Y}_c^T.
\end{equation}
As a result, we obtain $\widehat{\underline{\mathbf{s}}}_k^c(l^{'}) = f^{-1}(\widehat{\mathbf{S}}(k,l))\slash q^t_{s}(\hat{\tau}_k,\hat{\epsilon}_k)$ after compensating the phase rotation, where $ t = \mod[\frac{l-1}{S}]+1, s=\lceil l / S \rceil$. The mapping between $l$ and $l^{'}$ is determined by the interleaving pattern. Due to the sub-optimal estimation of offsets in Eq. \eqref{equ-43}, the residual errors will cause a slight phase rotation between $\widehat{\underline{\mathbf{s}}}_k^c$ and $\mathbf{\underline{\mathbf{s}}}_k^c$. We further introduce the variable $\rho$ to evaluate the residual phase rotations. For BPSK modulation \footnote{ We note that relatively low-order modulations are preferred to exploit the geometric characteristics of the constellation, and the algorithms can be easily extended to the QPSK case.}, it is given by
\begin{equation}
	\rho_{k} = \sum\nolimits_{\mathcal{R} \left\lbrace \widehat{\underline{\mathbf{s}}}_{k}^c \right\rbrace>0}\widehat{\underline{\mathbf{s}}}_{k}^c- \sum\nolimits_{\mathcal{R}\left\lbrace \widehat{\underline{\mathbf{s}}}_{k}^c \right\rbrace<0}{\widehat{\underline{\mathbf{s}}}_{k}^c}.
\end{equation}
Fig. \ref{pic-6} manifests the constellations of different phase rotation cases. The more precise offset estimation leads to the more sufficient compensation of the phase rotation, and thus the larger $\left\| \rho\right\|$. To this end, we generate a list of $N_s$ TO and CFO samples as coarse estimations by the GB-CR$^2$ algorithm. Correspondingly, the improved TO and CFO estimation can be obtained by
\begin{equation}
	\left[\hat{\tau}_k, \hat{\epsilon}_k \right] = \arg \max_{\hat{\tau}_{k,n}, \hat{\epsilon}_{k,n}} {\left\| \rho_{k,n}\right\|}, \forall n \in [1:N_s]. \label{equ-49}
\end{equation}
\par Furthermore, as we will see shortly in Section \ref{sec-VI}, an improved CE result can be obtained by plugging the enhanced  offsets into the GB-CR$^2$ algorithm and re-estimating the channel with known offsets. This is applied only to the flat fading channel case since TO errors have a greater impact on the CE accuracy than CFO errors. In the FSF channel case, TO is compensated during the CE process, thus without the need for the GB-CR$^2$ algorithm.  Once the channels are reconstructed and rotations are compensated, the subsequent LDPC decoding can be easily performed by the standard BP iterative structure, and thus omitted here. The overall algorithm is summarized in Alg. \ref{alg-2}. 

\par
	\begin{Remark}
		Note that the FSF and flat fading channels are distinct models and not contradictory as they apply to different scenarios: for a subcarrier spacing $\Delta f = 15$ kHz, when $S$ is small (e.g., $S=128$), it is considered as a narrowband system and the channel is regarded as flat fading. In contrast, for a larger $S$, such as $S=1024$, the system is wideband, and the channel is modeled as FSF. In this study, we consider both channel models and design two respective tailor-made CE methods. As will be demonstrated in Sec. \ref{sec-VI}, the proposed  CE methods, along with the GB-CR$^2$ algorithm, exhibit satisfactory performance in both channel models.
	\end{Remark}

\par  Note that the adoption of the ESOPs yields a significant benefit, i.e., it shifts the coupled phase rotations  from the codeword to the channel, although it is with the cost of limited codebook space due to orthogonality. Consequently, without requiring codebook quantization and expansion \cite{sun2022TWC}, the GB-CR$^2$ algorithm can readily address both TO and CFO by leveraging multi-stage channel information, while also resolve the codeword collisions due to the limited orthogonal space.

\section{Performance Analysis} \label{sec-5}

\subsection{Analysis of CE Performance}

\par In this subsection, we derive two benchmarks to evaluate the CE performance with the perfect knowledge of the channel profile. Let $\mathcal{I}$ denote the support-set and $\mathbf{G}_{O} = \mathbf{G}(:,\mathcal{I})$ represent the Oracle-sensing matrix comprising of the columns indexed by the support-set $\mathcal{I}$. Therefore, the Bayesian Fisher information matrix (FIM) is $\mathbf{J} = \mathbf{G}_{O}^H\boldsymbol{\Sigma}_{\mathbf{n}}^{-1}\mathbf{G}_{O}+\boldsymbol{\Sigma}_{\mathbf{x}}^{-1}$ \cite{BCRB}, where $\boldsymbol{\Sigma}_{\mathbf{x}}$ and $\boldsymbol{\Sigma}_{\mathbf{n}}$ denote the covariance matrices of $\mathbf{X}$ and $\mathbf{Z}$ in Eq. \eqref{equ-8}, respectively. Thus, the Oracle-MMSE estimation for $\mathbf{X}$ is given by
\begin{equation}
	\widehat{\mathbf{X}}^{\text{O-MMSE}} = \mathbf{J}^{-1}\mathbf{G}_{O}^{H}\boldsymbol{\Sigma}_{\mathbf{n}}^{-1}\mathbf{Y}. \label{equ-50}
\end{equation}
While the Oracle-BCRB is given by \cite{BCRB}
\begin{equation}
	\text{MSE}(\widehat{\mathbf{X}}) \geq M\text{Tr}(\mathbf{J}^{-1}), \label{equ-51}
\end{equation}
which serves as the lower bound on the MSE of CE. 

\subsection{Analysis of GB-CR$^2$ Algorithm}

\par In this subsection, we analyze the error probability of the GB-CR$^2$ algorithm, measured by the probabilities of missed detection (MD) and false alarm (FA), denoted by $\tilde{p}_{\mathrm{md}}$ and $\tilde{p}_{\mathrm{fa}}$, respectively. MD and FA denote that a valid path is miss-detected and an invalid path is falsely detected as valid, respectively. We define $\tilde{L}_{\mathrm{md}}$ and $\tilde{L}_{\mathrm{fa}}$ as the numbers of MD and FA paths of the GB-CR$^2$ algorithm, respectively, and $L^{(j)}$ is the number of EPs of the tree decoder at stage $j$. The expectation of $L^{(j)}$ is given by [\citen{Amal2020TIT}, Proposition 4]
\begin{equation}
	\mathbb{E}[L^{(j)}] = \sum\nolimits_{q=2}^j(K^{j-q}(K-1)\prod\nolimits_{l=q}^j 2^{-p_l}),  \label{equ-52}
\end{equation}
where $K$ is the number of nodes at each stage, and for simplicity we define $K \triangleq K_a$. Exploiting the path validity criterion of the GB-CR$^2$ algorithm, the expected number of EPs can be further reduced to $ \mathbb{E}[\tilde{L}_{\mathrm{fa}}] = (\delta^{'})^{\frac{T_p(T_p-1)}{2}}\mathbb{E}[L^{(T_p)}]$, where $\delta^{'}$ is the upper bound of the probability that two nodes on an invalid path satisfy Eq. \eqref{equ-34}, which follows that 
\begin{equation}
	\frac{\max\{\sigma_i^2, \sigma_j^2\}}{\sigma^2} < \frac{\Gamma^{-1}_{2SM}(\delta^{'}\slash 2)} 
{\Gamma^{-1}_{2SM}(1-\delta^{'}\slash 4)}, \label{equ-53}
\end{equation}
where $\sigma^2,\sigma_i^2$, and $\sigma_j^2$ are the same as those in Eq. \eqref{equ-36}, and the proof is simialr to Appendix \ref{app-1}, and thus omitted for brevity. Since the channels on the nodes of the invalid path belong to different users, statistically, $\sigma^2 = \sigma_i^2+\sigma_j^2$. In this way, a sufficiently small value of $\delta^{'}$ can fulfill Eq. \eqref{equ-53}, i.e., $\delta^{'}\ll 1$ in most cases. Thus, the number of EPs produced by the tree decoder can be further reduced by the GB-CR$^2$ algorithm, which will be verified by the numerical results in Sec. \ref{sec-VI}. Besides, provided that no error occurs in the activity detection, $\tilde{L}_{\mathrm{md}}$ follows the Binomial distribution, i.e., $\tilde{L}_{\mathrm{md}} \sim B(K_a,\delta^{\frac{T_p(T_p-1)}{2}})$, and $\mathbb{E}[\tilde{L}_{\mathrm{md}}]=\delta^{\frac{T_p(T_p-1)}{2}}K_a$, where $\delta$ satisfies Eq. \eqref{equ-35}. Therefore, based on the Markov inequality, $\tilde{p}_{\mathrm{md}}$ and $\tilde{p}_{\mathrm{fa}}$ are  bounded by
\begin{align}
	\tilde{p}_{\mathrm{md}} &= P(\tilde{L}_{\mathrm{md}} \geq 1) \leq \mathbb{E}[\tilde{L}_{\mathrm{md}}],\\
	\tilde{p}_{\mathrm{fa}} &= P(\tilde{L}_{\mathrm{fa}} \geq 1) \leq \mathbb{E}[\tilde{L}_{\mathrm{fa}}].
\end{align}

\subsection{Computational Complexity Analysis} 
\par We evaluate the computational complexity of the JADCE-MP-SBL algorithm by the required number of complex multiplications in each iteration. The computations for lines 5-9 and 10 in Alg. \ref{alg-1}  yield the complexities of $\mathcal{O}(NLSM)$ and $\mathcal{O}(SM)$, respectively. The computations related to the parameters $\epsilon, \gamma_n$, and $\lambda$ are with the complexities of $\mathcal{O}(NL), \mathcal{O}(M)$, and $\mathcal{O}(SM)$, respectively. For the MMSE estimation in the flat fading channel, the complexity is $\mathcal{O}(NM)$ for the unit orthogonal matrix $\mathbf{A}$. In general, the overall complexity order of the proposed algorithm is $\mathcal{O}(NLSM)$, linear with $N$ and $M$, making it computationally efficient for large codebooks and massive MIMO settings. 

\par Consequently, the complexity of the tree decoder is evaluated by the expected number of nodes for which parity checks must be computed \cite{Amal2020TIT}, which is given by
\begin{equation}
	\mathbb{E}[C_{\mathrm{tree}}]  = (T_p -1)K + \sum\nolimits_{j=2}^{T_p-1}\mathbb{E}[L^{(j)}]K.
\end{equation}
Besides, the complexity of the proposed GB-CR$^2$ algorithm for FSF and flat fading channels is $\mathcal{O}(MNQ)$ and $\mathcal{O}(MNDQ)$, respectively. Compared with the S-GAMP algorithm proposed in \cite{sun2022TWC} with a complexity of $\mathcal{O}(M^2ND^2Q^2)$, the complexity of the GB-CR$^2$ algorithm increases only linearly w.r.t. the quantization level, contributing to a more efficient approach for the offset estimation. 

\section{Numerical Results} \label{sec-VI}
\par In this section, we conduct numerical experiments to evaluate the performance of the proposed algorithms compared to existing works under both flat fading and FSF channel scenarios. The performance metrics are defined as follows: NMSE $ = 10\text{log}_{10} (\|\widehat{\mathbf{X}}-\mathbf{X}\|_F^2 \slash \|\mathbf{X} \|_F^2 )$, TO estimation error (TEE) $= \sum_{k}|\hat{\tau}_k-\tau_k|/K_a$, CFO estimation error (FEE) $= \sum_{k}|\hat{\epsilon}_k-\epsilon_k|/K_a$, block error rate (BLER)  $P_e = P_{\mathrm{md}} + P_{\mathrm{fa}}$, where $ P_{\mathrm{md}} $ and $ P_{\mathrm{fa}}$ are given by
\begin{align}  
	 P_{\mathrm{md}} &= \frac{1}{{{K_a}}}\sum\nolimits_{k \in {\mathcal{K}_a}} {P\left( {{\bm{v}_{{k}}} \notin \mathcal{L}} \right)},  \\
	 P_{\mathrm{fa}} &= \frac{{\left|  {\mathcal{L}\backslash \left\{ {{\bm{v}_{{k}}}:k \in {\mathcal{K}_a}} \right\}} \right|}}{{\left| \mathcal{L}\right|}}. \label{equ-56}
\end{align} 
In Eq. \eqref{equ-56}, $\mathcal{L}$ denotes the list of recovered messages. The system's signal-to-noise ratio (SNR) and energy-per-bit are defined as $\text{SNR} = 10\text{log}_{10}(\| \mathbf{GX}\|_F^2\slash{L_pM\sigma_n^2})$, and $E_b \slash N_0 = L_{\mathrm{tot}}P\slash BN_0$, respectively, where $L_{\mathrm{tot}}$ denotes the total channel uses and $P$ is the symbol power. The parameter settings are summarized in Tab. \ref{tab1}, which are used in the simulations unless otherwise specified.

 \begin{table*}[t]{}
 	\fontsize{8pt}{10pt}\selectfont
 	\renewcommand{\arraystretch}{1.3}
	\vspace{-0.2cm}
		\caption{Parameter Settings}
		\begin{center}
			\vspace{-0.2cm}	
			\begin{tabular}{| l | l || l | l |}
				\hline
				Parameters & Values & Parameters & Values \\
				\hline
				Number of subcarriers & $N_c = 2048$ &  Sub-block length & $J = 7$ \\
				\hline
				Subcarrier spacing & $\Delta f = 15$ kHz & Length of CP  & $L  = 72$  \\
				\hline
				Number of antennas & $M=16$ &  Quantization level & $D=9, Q=9$ \\
				\hline
				Number of channel paths & $L_k \in [5,15]$ & Maximum CFO & $\epsilon_{\max}=0.0133$\\
				\hline
				Number of OFDM symbols & $T_p = 4, T_c = 21$ & Message length & $B=100, B_p=14, B_c=86$   \\
				 \hline
 				\multirow {2} *{Subcarrier allocation} & ${\mathbf{s} = [1,2,\cdots,128]^T}^*$ & \multirow {2} *{Codeword length} & $L_p={128}^*, L_c = {2688}, L_{\mathrm{tot}} = 3200^*$\\
				 \cline{2-2} \cline{4-4} 
				 & ${\mathbf{s}=[1,2,\cdots,1024]^T}^{**}$ & & $L_p =  1024^{**}, L_c = 2688, L_{\mathrm{tot}} = 6874^{**}$  \\
				 \hline
			\end{tabular}
			\label{tab1}
	\end{center}
	\quad * and ** denote the flat fading and FSF channel scenarios, respectively.
\end{table*}

\subsection{Flat Fading Channel Scenario }

\par
\begin{figure}[htpb]
	\centerline{\includegraphics[width=0.48\textwidth]{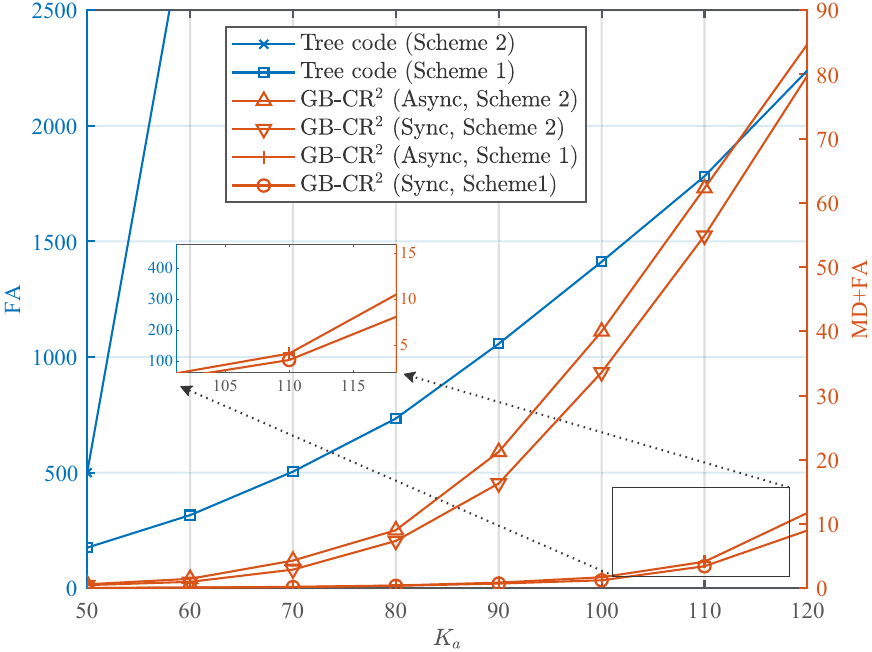}}
	\caption{The number of EPs (MD+FA) versus $K_a$.  }
	\label{pic-7}
\end{figure}

  In this subsection, we evaluate the performance of the proposed algorithm over the flat fading channels. Fig. \ref{pic-7} demonstrates the numbers of MD and FA paths of the GB-CR$^2$ algorithm versus the number $K_a$ of active users compared with the conventional tree code \cite{Amal2020TIT}  with $\text{SNR} = 6$ dB. Specifically, two parity bit allocation schemes are considered for the tree code: Scheme 1 under $[0,0,J,J]$ and Scheme 2 with $[0,0,0,J]$, resulting in $B_p = 14$ and $21$ in Schemes $1$ and $2$, respectively.  As depicted in Fig. \ref{pic-7}, for both coding schemes, the number of EPs of the GB-CR$^2$ algorithm is on the order of $10$, which is far less than that of the tree code, ranging from $10^2$ to $10^3$ in Scheme 1 and up to $10^5$ in Scheme 2. Furthermore, the GB-CR$^2$ algorithm exhibits only a slight performance degradation in the Async scenario compared to the Sync scenario, affirming the effectiveness of the proposed algorithm in validating paths and addressing offsets. In the following simulations, Scheme 1 is utilized in the coding process of  the preamble part. 

\par
\begin{figure*}[t] 
	\begin{minipage}{0.32\linewidth}
		\centerline{\includegraphics[width=\textwidth]{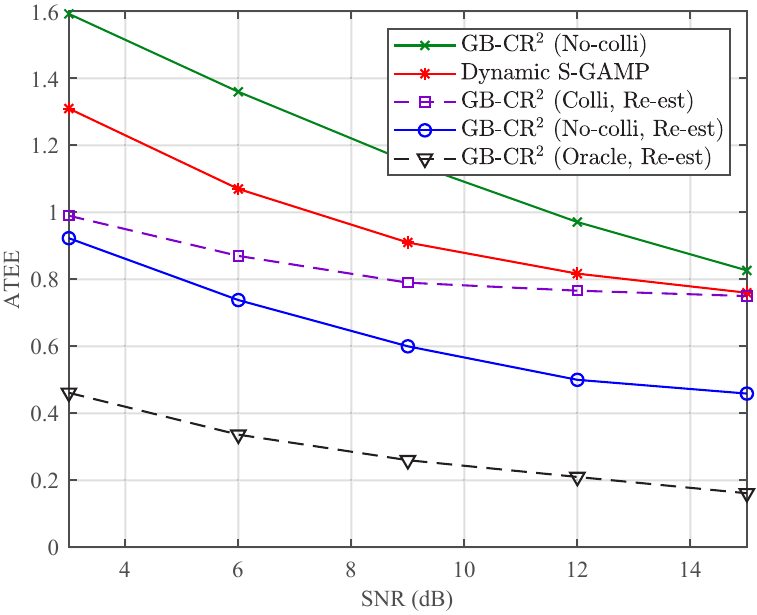}}
		\centerline{(a)}
	\end{minipage}
	\begin{minipage}{0.32\linewidth}
		\centerline{\includegraphics[width=\textwidth]{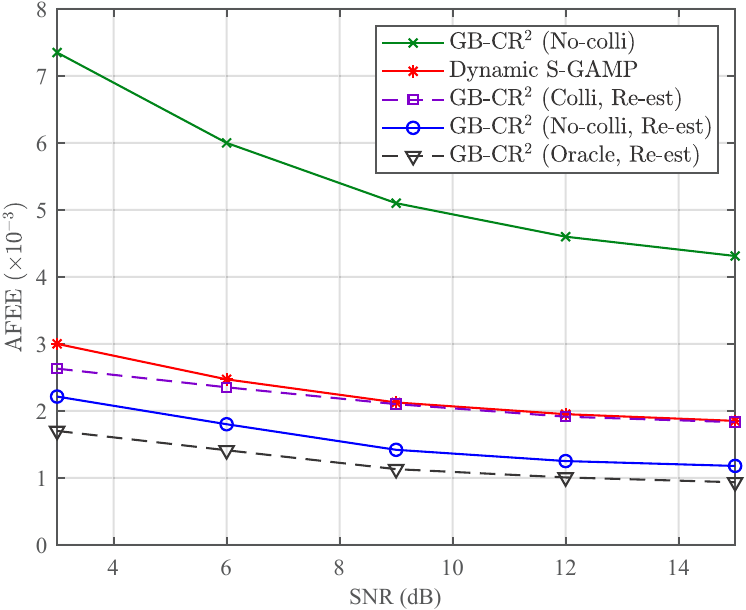}}
		\centerline{(b)}
	\end{minipage}
	\begin{minipage}{0.32\linewidth}
		\centerline{\includegraphics[width=\textwidth]{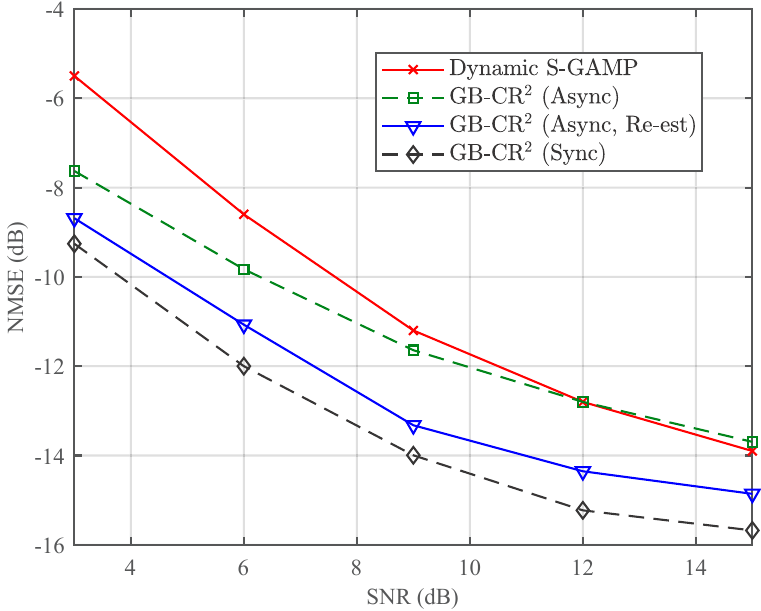}}
		\centerline{(c)}
	\end{minipage}
	\caption{Performance of the GB-CR$^2$ algorithm versus SNR. (a) The TO estimation performance comparison. (b) The CFO estimation performance comparison. (c) The CE performance comparison.}
	\label{pic-8}
\end{figure*}

 We further assess the offset and channel estimation performance of the GB-CR$^2$ algorithm compared with S-GAMP \cite{sun2022TWC}, where the codeword collision is not considered in S-GAMP due to the GF-RA scenario. In aligned with the work in \cite{sun2022TWC}, we set $M=4, K_a=8, B_p=8, J=4, L_p=18$, and $\mathbf{s} = [1,5,9,\cdots,69]^T$, resulting in a total of $72$ channel uses in the preamble part. The TEE and FEE performance versus SNR is depicted in Figs. \ref{pic-8}(a) and \ref{pic-8}(b), respectively, where ``Colli", ``No-colli", and ``Re-est" denote the collision and collision-free scenarios, and the constellation-aided offset correction scheme introduced in Sec \ref{sec-4}, respectively. And ``Oracle" refers to TO and CFO estimation under perfect decoding paths, i.e., GB-CR$^2$ is only tasked to estimate offsets based on the correct paths without resolving collisions and validating paths, which serves as the lower bound on the offset estimation performance. By leveraging the characteristics of the signal constellation in the coding part, the TO and CFO combination exhibiting the minimum phase rotation is chosen as the final estimation, which further enhances the offset estimation accuracy. Therefore, GB-CR$^2$ with ``Re-est" exhibits superiority in offset estimation over S-GAMP in both collision and collision-free scenarios. Particularly, in the latter case, the average TEE and FEE are reduced by up to $34.6 \%$ and $31.8 \%$, respectively. Nevertheless, influenced by the CE accuracy, the offset, especially the TO estimation, still shows a noticeable gap compared to the lower bound. Fig. \ref{pic-8}(c) presents the CE performance comparison of the involved algorithms versus SNR in the collision-free case. Similarly, by re-estimating the channel with corrected offsets, the proposed algorithm outperforms S-GAMP with an overall $2$ dB enhancement in the CE performance. Besides, the GB-CR$^2$ algorithm exhibits only $0.77$ dB performance loss in the Async case compared to the Sync case, revealing the efficacy of the proposed algorithm in handling offsets.

\par
\begin{figure}[htpb]
	\centerline{\includegraphics[width=0.45\textwidth]{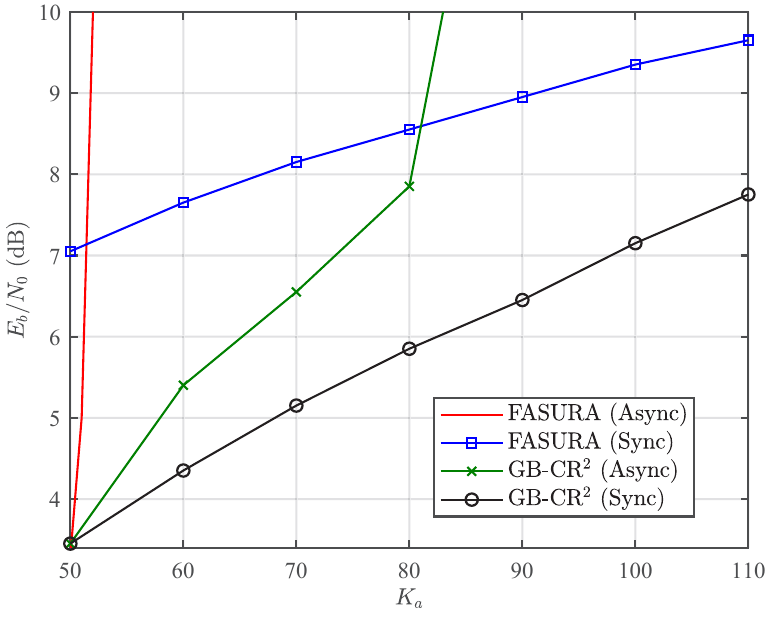}}
	\caption{The required $E_b\slash N_0$ for different schemes under the target $P_e \leq 0.001$ for the Sync case and $P_e \leq 0.01$ for the Async case.}
	\label{pic-9}
\end{figure}

 We then study the BLER performance of the proposed algorithm compared with existing works. Specifically, we set $J=7, B_p=14$ and adopt the identity matrix in the Async case, while the non-orthogonal Gaussian matrix is employed as the codebook in the Sync case with $J=12, B_p=24$. Fig. \ref{pic-9} depicts the required $E_b\slash N_0$ versus $K_a$ under the targeted $P_e$ for both the Sync and Async scenarios. It is shown that  GB-CR$^2$  significantly outperforms FASURA \cite{Gkag2023TCOM}, the current state-of-the-art scheme in URA, in both Sync and Async scenarios. Thanks to the remarkable offset compensation mechanism, our algorithm presents a satisfactory performance in the regime where $K_a \leq 80$ of the Async scenario, where FASURA fails to operate effectively. Moreover, in the Sync case, the proposed algorithm achieves a substantial energy saving of $2.7$ dB, while maintaining equivalent performance compared to its counterpart, which mainly benefits from the adopted IDMA framework for effectively reducing multi-user interference.

\begin{figure}[htpb]
	\centerline{\includegraphics[width=0.45\textwidth]{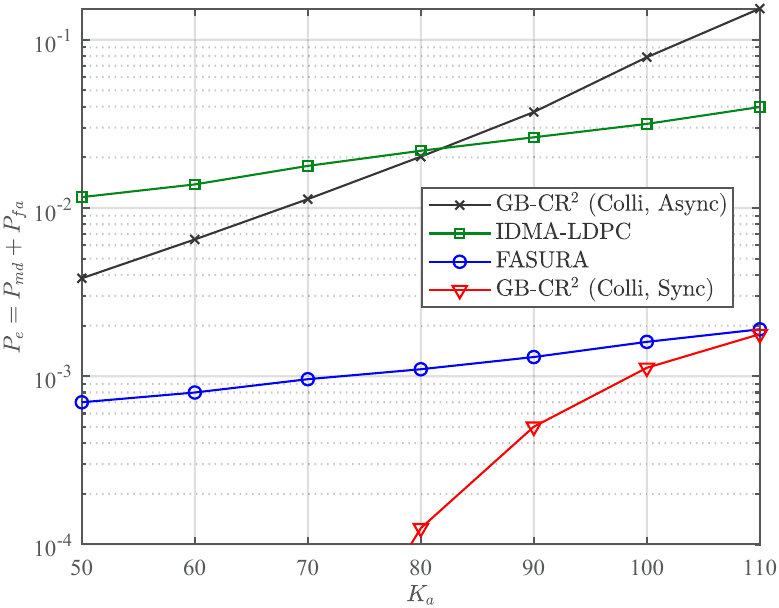}}
	\caption{The BLER performance comparisons versus $K_a$ with the fixed power $P=2$ dB.}
	\label{pic-10}
\end{figure}

\par Fig. \ref{pic-10} showcases the curve of $P_e$ as a function of $K_a$ with the transmitting power $P$ fixed to $2$ dB. To demonstrate the performance gain of the proposed collision resolution mechanism, we employ the two-phase transmission scheme recently introduced in \cite{Vem2019TCOM} as the baseline, denoted as IDMA-LDPC, where only one slot is transmitted in the preamble part, leading to unresolved collisions. We set $L_p=512$ and $B_p=12$ in IDMA-LDPC to ensure the identical channel uses. It is shown that the proposed algorithm significantly outperforms IDMA-LDPC by effectively resolving collisions. Moreover, GB-CR$^2$ exhibits a performance superiority to FASURA in the regime where $K_a \leq 110$. While FASURA encounters bottlenecks due to the absence of collision resolution, resulting in mediocre performance even with a small number of users.

\begin{figure}[htpb]
	\centerline{\includegraphics[width=0.45\textwidth]{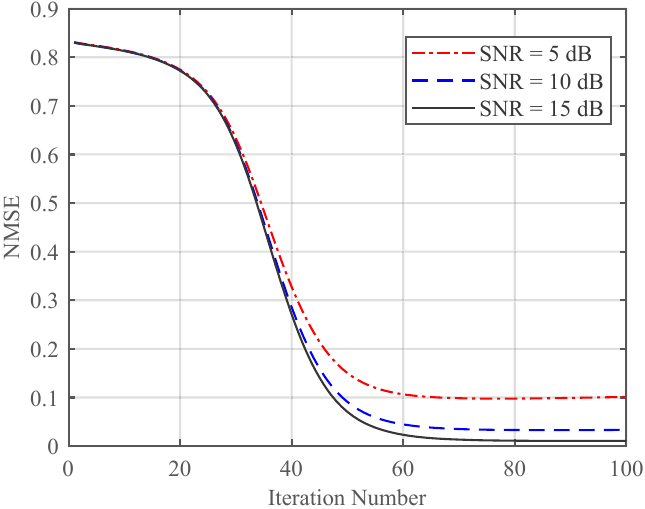}}
	\caption{ Convergence of the JADCE-MP-SBL algorithm.}
	\label{pic-11}
\end{figure}

\subsection{FSF Channel Scenario}

\par
\begin{figure}[htpb]
	\centerline{\includegraphics[width=0.45\textwidth]{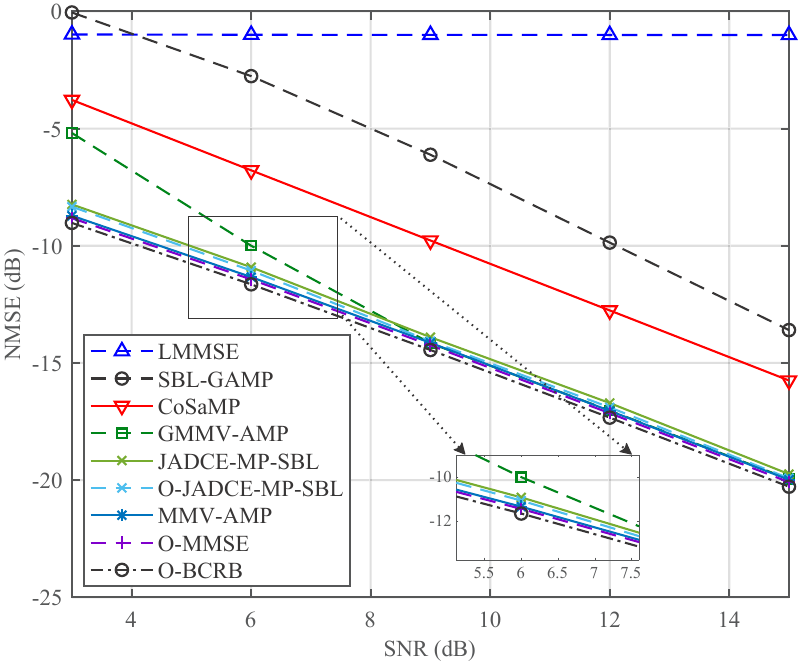}}
	\caption{The CE performance comparison with existing algorithms.}
	\label{pic-12}
\end{figure}

 In this subsection, we evaluate the CE and BLER performance of the GB-CR$^2$ algorithm over FSF channel scenarios. We consider a transmission bandwidth of $W=30$ MHz, the number $K_a$ of active users is $50$, and assume the multipath delay spread is $2 ~\mu$s \cite{URA_FS}. The number $L_k$ of multi-paths is fixed to $5$ for all active users in Figs. \ref{pic-11} and \ref{pic-12}. Fig. \ref{pic-11} illustrates the convergence of JADCE-MP-SBL under different SNRs, where the NMSE falls rapidly within the $20$-th to the $50$-th iterations, and converges with about $N_{\mathrm{iter}} = 80$, which is set to the maximum number of iterations of the algorithm in the subsequent simulations.

\par In Fig. \ref{pic-12}, we compare the CE performance of the proposed JADCE-MP-SBL algorithm with various benchmarks under the Sync scenario, where  ``O-MMSE" and ``O-BCRB" denote the lower bounds derived in Eqs. \eqref{equ-50} and \eqref{equ-51}, respectively, and ``O-JADCE-MP-SBL" refers to the Oracle JADCE-MP-SBL algorithm with known noise precision. Fig. \ref{pic-12} illustrates that the LMMSE estimator performs poorly due to the unknown supports of non-zero rows. The SBL-GAMP algorithm \cite{GAMP2018TSP}, based on the conventional SBL prior, which only considers the single element in the channel matrix, exhibits an overall performance loss of $7.5$ dB compared to the lower bound. Besides, the CoSaMP algorithm \cite{CoSaMP}, based on orthogonal matching pursuit, exhibits weak CE accuracy. Although MMV-AMP \cite{Liu2018tsp} and GMMV-AMP \cite{ke2020tps} demonstrate satisfactory CE performance and approach the lower bounds, they require prior information, such as activity probability and noise precision, which is difficult to obtain in practice. In contrast, the proposed JADCE-MP-SBL algorithm dynamically updates that information through the MP algorithm and achieves remarkable performance close to the lower bounds. Additionally, it incurs only a small performance loss compared to the case with known noise precision, i.e., O-JADCE-MP-SBL, demonstrating the superiority of the proposed algorithm in channel and prior information estimation.

\par
\begin{figure}[htpb]
	\centerline{\includegraphics[width=0.45\textwidth]{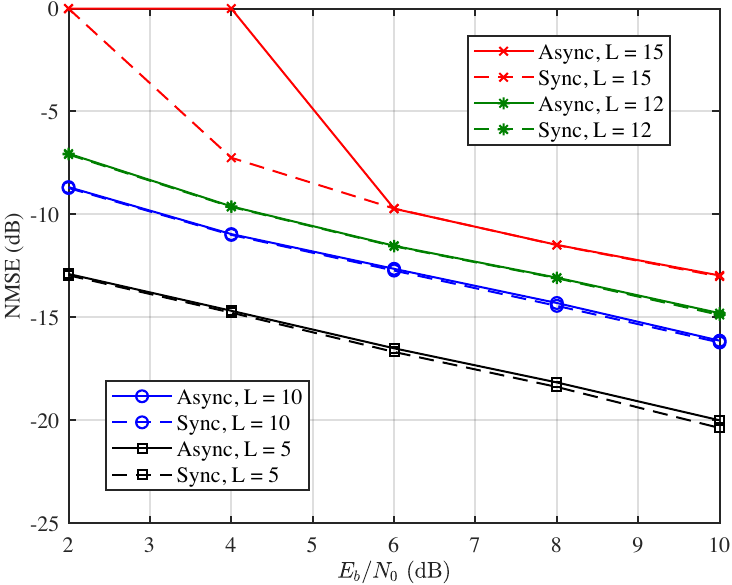}}
	\caption{The CE performance versus $E_b\slash N_0$ under different number of channel taps.}
	\label{pic-13}
\end{figure}

\begin{figure}[htpb]
	\centerline{\includegraphics[width=0.45\textwidth]{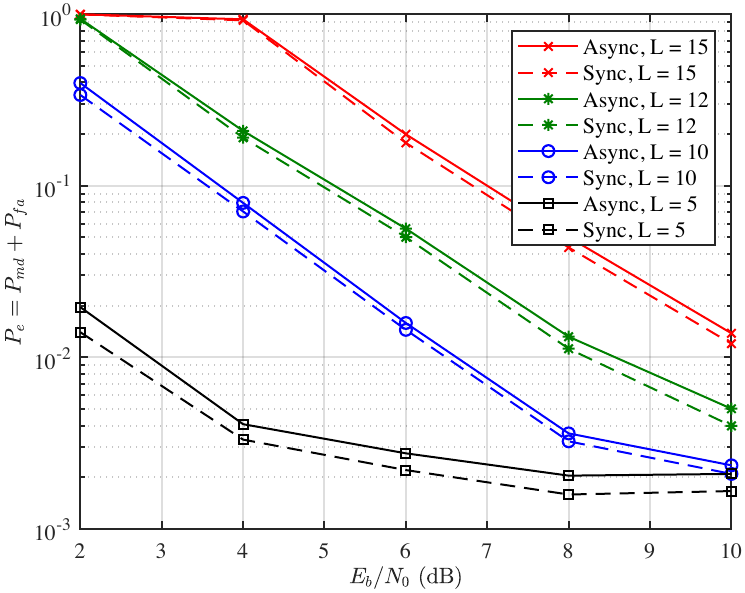}}
	\caption{The BLER performance versus $E_b\slash N_0$ under different number of channel taps.}
	\label{pic-14}
\end{figure}

 Finally, we evaluate the performance of the proposed algorithm with the number of channel taps $L_k \in \{5,10,12,15\}$. Fig. \ref{pic-13} demonstrates that the CE accuracy decreases as the number of multi-paths increases. This is because an increasing number of multi-paths reduces the sparsity of matrix $\mathbf{X}^t$ (cf. Fig. \ref{pic-4}), leading to a decline in CE accuracy as the number of channel responses to be estimated grows. This characteristic subsequently affects the performance of data decoding. In Fig. \ref{pic-14}, we evaluate the BLER performance of the proposed algorithm under the FSF channel scenario, where $J=12, B_p=24$. The vectors of the indexes of occupied subcarriers in the preamble and coding parts are $\mathbf{s}=[1,2,\cdots,1024]^T$  and $\mathbf{s}=[1,2,\cdots,128]^T$, respectively. Influenced by the CE results, the data decoding performance also decreases as the number of multi-paths increases. Besides, due to multi-user interference, the proposed algorithm encounters performance bottlenecks at $E_b\slash N_0=10$ dB for $L=5$ and $L=10$. Interestingly, after effectively addressing the CFO, the algorithm demonstrates comparable performance to that in the Sync scenario, highlighting its superior characteristics in Async scenarios.

\section{Conclusion} \label{sec-7}
\par This paper considered the Async MIMO-OFDM URA system over the FSF channel with the existence of both TO and CFO and non-negligible codeword collisions. By leveraging the dual sparsity of the channel in both CD and DD, we established the JADCE-MP-SBL algorithm combined with BP and MF to iteratively retrieve the FSF channels. Furthermore, we proposed a novel algorithmic solution called GB-CR$^2$ to reconstruct superimposed channels and compensate for TO and CFO with linear complexity w.r.t. the quantization level. In addition, we applied the proposed algorithm to the flat fading channel case, where the performance was further improved by leveraging the geometric characteristics of signal constellations. Numerical results manifested the reliability of the presented algorithms with superior performance while maintaining reduced complexity.

 \appendix
 \subsection{Proof of Proposition 1}\label{app-1}
 \par  We first define $\mathbf{H}_1, \mathbf{H}_2 \in \mathbb{C}^{K\times M}$, which are the channels of the same user estimated on different slots, and the entries in each are i.i.d. following $\mathcal{CN}(0,\sigma_1^2)$ and $\mathcal{CN}(0,\sigma_2^2)$, respectively. Denote $\mathbf{H} \triangleq \mathbf{H}_1- \mathbf{H}_2$ and the entries follow $\mathcal{CN}(0,\sigma^2)$, where the entries in $\mathbf{H}$ are i.i.d. since the channel components belonging to the same user in $\mathbf{H}_1$ and $\mathbf{H}_2$ are eliminated and only non-correlated components remain. We have
 \begin{equation}
 	\begin{aligned}
 		 	&P\{ \| \mathbf{H} \|_F^2 > \max \{\| \mathbf{H}_1\|_F^2, \| \mathbf{H}_2 \|_F^2\} \}  \\
 		&\leq P \{ \| \mathbf{H}\|_F^2 > \alpha \} + P \{\max \{\| \mathbf{H}_1\|_F^2, \| \mathbf{H}_2 \|_F^2\} < \alpha\},
 	\end{aligned}  \label{equ-app-1}
 \end{equation}
where $\alpha$ is a constant. Note that $ \frac{2}{\sigma^2}  \| \mathbf{H}\|_F^2 \sim \chi_{2KM}^2$ and we consider an upper bound of $P \{ \| \mathbf{H}\|_F^2 > \alpha \} $ such that
\begin{equation}
	P( \| \mathbf{H}\|_F^2>\alpha) = 1- \Gamma_{2KM}({2\alpha}\slash{\sigma^2} ) < \delta \slash 2   \label{equ-app-2}
\end{equation}
in the case of 
\begin{equation}
		\alpha > \frac{\sigma^2}{2}  \Gamma_{2KM}^{-1}(1-\delta \slash 2).  \label{equ-app-3}
\end{equation}
 Due to the possible correlation between $\mathbf{H}_1$ and $\mathbf{H}_2$, the upper bound for $P \{\max \{\| \mathbf{H}_1\|_F^2, \| \mathbf{H}_2 \|_F^2\} < \alpha\}$ is obtained as
\begin{equation}
	\begin{aligned}
		& P \{\max \{\| \mathbf{H}_1\|_F^2, \| \mathbf{H}_2 \|_F^2\} < \alpha\} \\
		& \leq \min \{\Gamma_{2KM}({2\alpha}\slash{\sigma_1^2} ),\Gamma_{2KM}({2\alpha}\slash{\sigma_2^2})\} < \delta\slash 2
	\end{aligned}  \label{equ-app-4}
\end{equation}
when
\begin{equation}
	\alpha < \max \{ \frac{\sigma_1^2}{2} \Gamma_{2KM}^{-1} (\delta\slash 2) , \frac{\sigma_2^2}{2} \Gamma_{2KM}^{-1} (\delta\slash 2) \}.  \label{equ-app-5}
\end{equation}
Therefore, to make Eqs. \eqref{equ-app-2} and \eqref{equ-app-4} both hold, we have
\begin{equation}
	 \frac{\sigma^2}{2}  \Gamma_{2KM}^{-1}(1-\delta \slash 2) < \max \{{\sigma_1^2}\slash{2}, {\sigma_2^2}\slash{2}\} \Gamma_{2KM}^{-1} (\delta\slash 2). 
\end{equation}
After some algebra, Eq. \eqref{equ-36} is obtained. Therefore, we have
\begin{equation}
	P\{ \| \mathbf{H}_1-\mathbf{H}_2 \|_F^2 > \max \{\| \mathbf{H}_1\|_F^2, \| \mathbf{H}_2 \|_F^2\} \} < \delta. \label{equ-app-6}
\end{equation}
Thus, Proposition \ref{prop-1} is proved.

\subsection{Proof of NP Hypothesis Testing} \label{app-2}
\par Based on Eq. \eqref{equ-37}, we define $\mathbf{X} \triangleq \widehat{\underline{\mathbf{X}}}_i- \widehat{\underline{\mathbf{X}}}_j \in \mathbb{C}^{K\times M}$ and establish the following binary hypothesis testing problem:
\begin{equation}
	\begin{aligned}
		 \mathbf{x}_k| \mathcal{H}_0 \sim \mathcal{CN}(0,\sigma_0^2 \mathbf{I}_M), \\
		\mathbf{x}_k | \mathcal{H}_1 \sim \mathcal{CN}(0,\sigma_1^2 \mathbf{I}_M),
	\end{aligned} \label{equ-app-2-1}
\end{equation}
where $\sigma_0^2 = 2\hat{\sigma}_e^2, \sigma_1^2 = 2\hat{\sigma}_e^2  + n\sigma_h^2$, $\mathbf{x}_k$ is the $k$-th row of $\mathbf{X}$, $\mathcal{H}_0$ and $\mathcal{H}_1$ are null and alternative hypotheses corresponding to non-collided and collided situations, respectively. $\sigma_h^2$ is channel's power  and $n$ is the number of collided nodes. Thus, the likelihood ratio for Eq. \eqref{equ-app-2-1} is given by 
\begin{equation}
	L(\mathbf{X}) = \frac{p(\mathbf{X} | \mathcal{H}_1)}{p(\mathbf{X} | \mathcal{H}_0)} \overset{(a)}{=} \left( {\sigma_0}\slash{\sigma_1} \right)^{2KM} \exp \left\lbrace  {\|\mathbf{X}\|_F^2 \slash \sigma^2} \right\rbrace  \label{equ-app-2-2}
\end{equation}
where $(a)$ holds because $p(\mathbf{X})$ = $\prod\nolimits_{k} p(\mathbf{x}_k)$ and $\sigma^2 = {\sigma_0^2\sigma_1^2}\slash {(\sigma_1^2- \sigma_0^2)}$. Thus, the NP test for the decision of collision is
\begin{equation}
	\delta_{NP} (\mathbf{X})  = \left\lbrace \begin{array}{l}
		1, ~~ L(\mathbf{X}) \geq \gamma_0, \\
		0, ~~  L(\mathbf{X}) < \gamma_0
	\end{array} \right. 
	=  \left\lbrace \begin{array}{l}
		1, ~~  \|\mathbf{X}\|_F^2 \geq \gamma, \\
		0, ~~  \|\mathbf{X}\|_F^2 < \gamma,
	\end{array} \right.   \label{equ-app-2-3}
\end{equation}
where $\gamma = 2KM \sigma^2 \ln ( \gamma_0 \sigma_1 \slash \sigma_0)$. The FA probability of the decision is obtained by
\begin{equation}
	P_{\mathrm{fa}}(\delta_{NP}) = P(\|\mathbf{X}\|_F^2 \geq \gamma | \mathcal{H}_0) \overset{(a)}{=} 1- \Gamma_{2KM}(\gamma\slash \hat{\sigma}_e^2 ),   \label{equ-app-2-4}
\end{equation}
where $(a)$ follows from the fact that $\|\mathbf{X}\|_F^2 | \mathcal{H}_0\sim\hat{\sigma}_e^2  \chi^2_{2KM}$. Under the $\zeta$-level NP test, it should satisfy $P_{\mathrm{fa}}(\delta_{\mathrm{NP}}) \leq \zeta$. Thus, the threshold to verify the collision nodes is given by
\begin{equation}
	\gamma =  \hat{\sigma}_e^2 \Gamma_{2KM}^{-1}(1-\zeta).
\end{equation}
Thus, the proof is completed.

\end{document}